\documentclass[runningheads]{llncs}
\usepackage{graphicx}
\usepackage{graphics}
\usepackage{csquotes}
\PassOptionsToPackage{hyphens}{url}\usepackage{hyperref}
\usepackage{listings}
\usepackage[space]{cite}
\usepackage{colortbl}
\usepackage{booktabs}

\definecolor{Gray}{rgb}{0.99, 0.88, 0.88}
\usepackage[numbers,sort&compress]{natbib}

\lstdefinelanguage{JavaScript}{
	keywords={},
	keywordstyle=\color{bluegrey}\bfseries,
	morekeywords=[2]{attributes, class, classend, do, empty, endif, endwhile, fail, function, functionend, if, implements, in, inherit, inout, not, of, operations, out, return, set, then, types, while, use},
	keywordstyle={[2]\color{violet}\bfseries},
	otherkeywords={@param, @returns, @author, @type, @link, @see},
	sensitive=false,
	morecomment=[l][\color{lstgreen}]{//},
	morecomment=[s][\color{lstgreen}]{/*}{*/},
	morecomment=[s][\color{javadoc}]{/**}{*/},
	morestring=[b]',
	morestring=[b]"
}

\makeatletter
\newcommand{\printfnsymbol}[1]{%
  \textsuperscript{\@fnsymbol{#1}}%
}
\usepackage{hyperref}[bookmarks=true,bookmarksnumbered=true]
\makeatother
\begin{document}

%
%
\title{Addressing the Regulatory Gap: Moving Towards an EU AI Audit Ecosystem Beyond the AIA by Including Civil Society}
\titlerunning{Towards an EU AI Auditing Ecosystem}
%
\author{
David Hartmann\inst{1,2}\orcidID{0000-0001-9745-5287}, José Renato Laranjeira de Pereira\inst{3, 6}\orcidID{0000-0002-9605-8121}, Chiara Streitbörger\inst{4}, \and Bettina Berendt\inst{1,2,5}\orcidID{0000-0002-8003-3413}\\
}

%
\authorrunning{D. Hartmann et al.}
%

\institute{Faculty of Electrical Engineering and Computer Science, TU Berlin, Berlin, Germany\\
\and
Weizenbaum Institute for the Networked Society, Berlin, Germany\\
\and
Laboratório de Políticas Públicas e Internet - LAPIN, Brasília, DF, Brazil\\
\and 
Kammergericht Berlin, Berlin, Germany\\
\and 
Department of Computer Science, KU Leuven, Leuven, Belgium\\
\and
Bonn Sustainable AI Lab, University of Bonn, Bonn, Germany\\
}
\maketitle              
\begin{abstract}
The European legislature has proposed the Digital Services Act (DSA) and Artificial Intelligence Act (AIA) to regulate platforms and Artificial Intelligence (AI) products. We review to what extent third-party audits are part of both laws and how is access to information on models and the data provided. By considering the value of third-party audits and third-party data access in an audit ecosystem, we identify a regulatory gap in that the AIA does not provide access to data for researchers and civil society. Our contributions to the literature include: (1) Defining an AI audit ecosystem incorporating compliance and oversight. (2) Highlighting a regulatory gap within the DSA and AIA regulatory framework, preventing the establishment of an AI audit ecosystem that has effective oversight by civil society and academia. (3) Emphasizing that third-party audits by research and civil society must be part of that ecosystem, we call for AIA amendments and delegated acts to include data and model access for certain AI products. Furthermore, we call for the DSA to provide NGOs and investigative journalists with data access to platforms by delegated acts and for adaptions and amendments of the AIA to provide third-party audits and data and model access, at least for high-risk systems, to at least reduce the regulatory gap that exists in the EU. Regulations modeled after the bloc's AI regulations should enable data access and third-party audits, fostering an AI audit ecosystem that promotes compliance and oversight mechanisms.

\keywords{Auditing, Accountability, Data Access, AI Act, DSA, Black-box Audits, AI}
\end{abstract}
\section{Introduction}
Artificial intelligence (AI) technologies can be found in technologies and products such as machine learning (ML)-based automated decision-making (ADM) systems, such as social media recommendation systems, and general-purpose AI, which includes generative AI models. These technologies have been proven to cause severe harm to individuals, groups, and society, especially through their capacity to enable discriminatory practices and for their opaqueness, such as in Rotterdam's welfare fraud, whereby ADM systems deployed were biased against immigrants, who were frequently labeled as fraudsters against the system \cite{wire}. Scholars have also argued that YouTube's algorithmic-based recommender systems \cite{ribeiro20202} were more prone to recommend extreme content on the platform, helping spread disinformation and harmful content. Furthermore, harmful biases have been uncovered in ML models such as in the generative AI Stable Diffusion \cite{bloomberg}, face recognition APIs \cite{buolamwini_gender_nodate}, and ChatGPT \cite{ray_chatgpt_2023}. 

The European legislature has responded with different regulatory approaches to address these risks. For example, it has proposed regulating very large online platforms \footnote{Not restricted to AI-driven algorithms} (VLOPs) and very large online search engines (VLOSEs) via the Digital Services Act (DSA) \cite{dsa}, as well as AI-driven products, services, and systems in the territory of the European Union (EU) via the Artificial Intelligence Act (AIA) \cite{AIAct2024}. 

Among other regulatory tools, both the DSA and the AI Act have provisions related to systematic evaluations of AI systems -- denoted by the term audit -- by agents that are both internal or external to the organisation developing or deploying these technologies and which aim at creating public accountability regarding these machine's behavior \cite{raji_outsider_2022,bandy_problematic_2021, raji_actionable_2019}. These provisions include self-assessments both before deployment of the system — as is the case in the AI Act for ``high-risk'' AI systems which have to undergo a conformity assessment prior to be place in the EU Market \cite{edwards2022} — and after deployment — as are the annual independent audits that VLOPs and VLOSEs have to be subjected to under Article 37 of the DSA \cite{dsa-audit}. 

It has been argued by \citet{mokander_conformity_2022} that the EU regulation thereby sketches a \textit{de facto} EU-wide ecosystem for auditing AI systems. However, \citet{edwards2022} has argued that self-assessments before deployment and post-market monitoring -- we go with \citet{mokander_conformity_2022} to frame this as a form of internal auditing -- do not provide sufficient oversight. Therefore, \citet{edwards2022} has raised the question of the extent to which audits and assessments by third parties (regulator, research, and civil society) are necessary for sufficient oversight. 

As this paper will demonstrate, access to elements such as the model and training data of the algorithmic system is crucial to facilitate external auditing. In this sense, we reiterate and strengthen, on the one hand, the argument of \citet{casper_black-box_2024} that third-party black-box access —  in which ``auditors can only query the system and observe its outputs'', as opposed to ``white-box access to the system's inner workings'' — is insufficient and, on the other, AlgorithmWatch's concern that the fact that third-party access is lacking in the AIA is worrisome for the efficacy of the regulation \cite{Algorithmwatch2021}. Moreover, while the DSA includes third-party access by vetted researchers, scholars have argued that companies could leverage their market power against external auditors (i.e. audit capture) \cite{laux_taming_2021}. This is why we call for extended access for civil society and journalists to strengthen auditors positions and oversight by civil society during the implementation of both the DSA and for future regulations.

In general, researchers wonder “who will audit internal auditors?” \cite{costanza-chock_who_2022, albert_john_diverse_2023, edwards2022}. Therefore, we analyze whether the DSA and the AIA can create an effective, diverse EU audit ecosystem that is capable to protect the rights that they aim to protect. To do so, we examine the importance of third-party audits by researchers and civil society. We find that third-party audits are essential for the establishment of an AI auditing ecosystem in the EU that ensures accountability by compliance and oversight. We argue that the inclusion of such audits must be strengthened by considering the existing provisions of the DSA and the provisions of the AIA, which comes into force in 2024. The question of how audits conducted by the regulator should be included in an AI auditing ecosystem unfortunately goes beyond the scope of the paper, although researchers have justifiably insisted on the importance of this issue (see e.g. \cite{heuer_audit_2021}). 

We contribute to the existing literature concerning third-party audits in three ways: (1) We define an AI audit ecosystem that accounts for compliance and oversight. (2) We demonstrate the existence of a regulatory gap as the DSA and AIA will not lead to a diverse AI audit ecosystem that includes civil society and, in the case of the AIA, even vetted researchers. (3) We emphasize that external audits by vetted, independent research centres, journalists and civil society organisations have to be part of that ecosystem. For this reason, we argue that the AI Act has lost an important opportunity to include should include broader access to AI systems, as it is provided in the case of the DSA, and that this gap should be addressed by future regulation and AIA amendments.

First, we introduce the terms related to algorithmic audits and the different types of audits (section \ref{sec:terminology}). We then analyze the current regulatory framework on third-party audits (section \ref{sec:framework}). The following section addresses third-party audit case studies and defines an AI audit ecosystem that ensures compliance and oversight. Furthermore, we show that third-party audits by researchers and civil society are vital for an AI audit ecosystem (section \ref{sec:ecosystem}). Finally, we argue that the current EU legislation -- the DSA and AIA in combination -- will not be sufficient to create a diverse AI audit ecosystem that includes researchers and civil society representatives. We then make concrete proposals for future regulation, as well as for delegated acts and potential amendments for the DSA and the AIA (section \ref{sec:gap}). 
\section{Algorithm Audits: Terminology and Key Properties}
\label{sec:terminology}
In this section, our primary objective is to establish a robust foundation of terminology and key attributes inherent to algorithm audits. Consequently, we will distinguish audit types, examine the scope of audit types, and underline the role of sociotechnical system thinking in audits. These examinations are conducted in preparation for the analysis of the role of third parties in the DSA and AIA. Additionally, we will revisit these audit types in the discussion of creating a diverse AI audit ecosystem.

The term “audit” \footnote{\citet{sandvig_auditing_2014} have noted that ``although the word 'audit' may evoke financial accounting, the original audit studies were developed by government economists to detect racial discrimination in housing.'' Later, they were introduced in the social science as field experiments to test for discriminatory behavior \cite{gaddis_introduction_2017}.} is used in various ways. For algorithm audits\footnote{In the following, we will use the terms "audit" and "algorithm audit" interchangeably.}, some (e.g. \citet{koshiyama_towards_2021}) have provided a more technical definition of an algorithm audit, which is related to verification and compliance, and others  (e.g. \citet{sandvig_auditing_2014} and \citet{HCI-083}) have used the term to describe a specific but still systematic targeted test of a particular aspect of a system (e.g., bias), originating from social science ``audit studies.'' 

Two examples demonstrate the various ways that the term has been used. Firstly, \citet{bandy_problematic_2021} identifies a research gap in audits of Twitter. Twitter is the most researched social media platform due to its formerly openly-accessible API \cite{olteanu2019social}. This suggests a conceptual problem: What differentiates research on Twitter from an audit conducted on Twitter? In the second example, \citet{HCI-083} defines audits as systematic external evaluations by third parties, serving as a form of activism with internal knowledge of the process or system being studied.

We base our understanding of audits on \citet{raji_outsider_2022} and \citet{bandy_problematic_2021} by defining ``audit'' as an empirical study that investigates algorithmic systems and evaluates performance relative to expected behavior as part of a broader accountability process. The focus of an audit is to examine for potential problematic behavior, which refers to system behavior with the potential to cause harm to individuals, groups, or society \cite{bandy_problematic_2021}. ``Empirical studies'' refers to qualitative and quantitative studies, as system expectations can be articulated and assessed in either mode \cite{raji_outsider_2022}. We note that this definition is also agnostic to the time of the audit (before or after the deployment).

In a systematic review of algorithm audits, \citet{bandy_problematic_2021} points out that racial discrimination remains a central focus. Thus, Bias audits address algorithmic discrimination based on race and other protected attributes, such as age, gender, socioeconomic status, religion, and intersectional identities. However, apart from representational and allocative harms related to discrimination, AI systems can pose other potential risks. Therefore, audits may need to have different scopes to address quality of service harms, interpersonal harms, and social system harms \cite{shelby_sociotechnical_2022}. To address these harm types, audits such as robustness audits, interpretability and explainability audits, security audits, and privacy audits can be conducted \cite{koshiyama_towards_2021}.

Based on \citet{institute_algorithmic_2023}, the following definitions are used in this paper:
\begin{itemize}
    \item first-party audits: internal audits which are conducted within an organization that developed the AI system or product based on specific metrics, toolkits, and requirements.  
    
    \item second-party audits: conducted by contractors with the developer (audits-as-a-service), including consulting companies as the ``big four,'' as well as companies that specialize in audits. While maintaining a degree of independence from deployers, risk ``audit washing" (see section \ref{sec:washing}) by potentially catering to their clients' interests.
    
    \item third-party audits: are conducted by independent organizations which have no contractual relationship to the developing organization. Such organizations include independent researchers, journalists, law firms, regulators, civil society, non-governmental organizations (NGOs), users, and affected communities.
\end{itemize} 
We adopt the audit design considerations that \citet{raji_outsider_2022} has developed by analyzing various audit studies to account for the range of possible algorithmic audits. These considerations include (1) target identification and audit scope; (2) auditor independence; (3) auditor privileges; (4) auditor professionalization and conduct standards; and (5) when to audit and post-audit actions. Given these considerations, five audit types can be distinguished: algorithmic risk assessment, conformity assessments, research and civil society audits, regulatory inspection and certification, and sociotechnical audits. The characteristics of these audits and their assignment to first-party, second-party, and third-party audits can be found in Table \ref{tab:characteristica}.

\begin{table}[h!]
    \centering
    \resizebox{\columnwidth}{!}{
    \begin{tabular}{p{2cm}||p{2.5cm}|p{2.5cm}|p{2.5cm}|p{2.5cm}|p{2.5cm}|p{2.5cm}|}
    \cline{2-7}
     &\multicolumn{3}{c|}{First-party assessments and audits} &\multicolumn{3}{c|}{Second- and Third-party assessments and audits}\\ 
     \cline{2-7}\noalign{\vskip\doublerulesep
         \vskip-\arrayrulewidth}\cline{2-7}
     &Algorithmic risk assessment, algorithmic impact assessment& Ethic-based auditing and AI safety auditing&Internal conformity assessment and post-market monitoring&Research and civil society algorithm audits, bias audits, adversarial audits& Regulatory inspection, external conformity assessments and certification&Algorithmic impact evaluations, sociotechnical audits, ecosystem audits \\ \hline\noalign{\vskip\doublerulesep
         \vskip-\arrayrulewidth}
    
     \rowcolor{Gray}
     Example(s) & Case study algorithmic impact assessment for data access in a healthcare context \cite{groves_algorithmic_nodate}, data protection impact assessment \cite{bieker_process_2016}& End-to-end framework for internal algorithmic
auditing by \citet{raji2020closing}& AIA conformity assessments, example procedure by \cite{floridi_capai_2022}; Medical Device Regulation Application Procedure \cite{doi:10.1177/2168479017716712} & Gender Shades \cite{buolamwini_gender_nodate}, Audit of the Rotterdams' welfare fraud ADM \cite{wire,noauthor_suspicion_nodate}, Audit Austrian AMS algorithm \cite{allhutter2020ams}& UK Information Commissioner’s Office’s ‘Guidance on the AI auditing framework'\cite{kazim2021ai}& Stanford’s Impact
evaluation of a predictive risk modeling tool \cite{goldhaber2019impact}, sociotechnical audit that assess police use of facial recognition\cite{sociotechnical}\\
     Auditors & Creators or
commissioners
of the algorithmic
system, contractors such as consulting companies& Creators or
commissioners
of the algorithmic
system, contractors such as consulting companies &Creators or
commissioners
of the algorithmic
system, contractors such as consulting companies &Researchers, investigative journalists, data scientists, NGOs, affected communities, or other stakeholders & Regulator or other auditing and compliance
professionals that have an institutional role&Researchers,
policymakers, investigative journalists, data scientists, NGOs, affected communities, or other stakeholders\\
 \rowcolor{Gray}
     Target Identification \& Audit Scope & Broad scope: focus on the risks and negative consequences of the deployment of an algorithmic system & Narrow scope: focus predominantly on technical prevention and uncovering of specific harms (e.g. bias audit, safety audit)&Broad scope: focus on
an algorithmic
system’s
compliance with
regulation and mostly non-technical and process-oriented &Narrow scope: focus predominantly on technical prevention and uncovering of specific harms (e.g. bias audit)& Broad scope: focus on
an algorithmic
system’s
compliance with
regulation& Broad scope: assessing possible
societal impacts
of an algorithmic
system on the
users or population
it affects after it
is in use\\
    
    Auditor Independence& Internal and thus no independence or contractual dependence&Internal and thus no independence or contractual dependence& Internal and thus no independence or contractual dependence&External by independent organizations with no obligation&External by independent institution, non-genuine independence when auditing administrative state-run ADMs& External by independent organizations with no
obligation\\
\rowcolor{Gray}
    Auditor Professionalization & Formal: done by deployers, internal teams or accredited second-parties& Formal: Done by deployers, internal teams or accredited second-parties& Formal: done by deployers, internal teams or accredited second-parties&Mostly informal: accreditation and cooperation with company possible , although basic access could be possible without accreditation ('right to scrape')&Formal: Accreditation&Mostly informal: accreditation and cooperation with company possible (vetted researchers in DSA)\\

    When to audit & Before deployment &Before deployment& Before deployment, post-market monitoring is possible &After deployment& After or before deployment, potentially an ongoing process&After deployment\\
     \hline
    \end{tabular}
    }
    \caption{This table is inspired by and derived from \cite{ada2020examining,HCI-083,mokander_ethics-based_2021,raji_outsider_2022} and should give an overview of assessments and audits that we surveyed. The considerations of \citet{raji_outsider_2022} were used to display the differences of these assessment and audit types. However, this table claims neither completeness nor complete adequacy, as this topic is a relatively emerging one with respect to algorithmic systems. At the same time, a systematic presentation of assessments and audits differences is not the focus of this paper, even though we conceptually argue and advocate a broad deployment of assessments and audits in the AI lifecycle within the AI audit ecosystem.}
    \label{tab:characteristica}
    
\end{table}


Algorithms cannot be divorced from the contexts in which they are applied \cite{veale_governing_2019} and represent a complex network containing economic, environmental, political, and social interests and outcomes affecting their functioning. The entanglement of technical components and social elements leads to the typical acknowledgment of AI systems as sociotechnical systems\footnote{Due to a broader understanding of their ecological impacts in recent years, \cite{rakova_algorithms_2023} has added an “ecological” element to this notion, meaning that AI systems are socioecological-technical systems. Thus, it may be necessary to include ecological harms that AI causes in audits.} \cite{sartori2022sociotechnical} And, despite the common perception of audits as predominantly technical (see e.g. \citet{koshiyama_towards_2021}), internally conducted impact assessment as well as externally conducted impact evaluations and sociotechnical audits encompass a broader spectrum. They evaluate the impact of an algorithmic system on a population in a comprehensive framework that encompasses technical elements -- most of the time limited technical elements -- and include the social elements of the system \cite{ada2020examining}. We will come back to the assessment of sociotechnical systems in section \ref{sec:humanrights}.

Ethics-based or AI safety internal auditing and adversarial audits by researchers and civil society have a more technical scope and tend to focus on specific harms, e.g. bias audits. Nevertheless, drawing a clear line of demarcation between narrow and broad scope of audit is difficult. While maintaining a however central focus on sociotechnical systems, it is essential to consider the data that substantiate claims concerning specific harms attributed to an AI system. An example of this difficult distinction is that \citet{stahl2023systematic} includes bias mitigation assessments in the systematic review on impact assessments, which we have introduced as ethics-based audits or bias audits. Thus, these kind of audits could potentially be part of an impact assessment. 

Conformity assessments \cite{mokander_conformity_2022} and regulatory inspection \cite{ada2020examining} are systematic tests that evaluate if a system is compliant with a specification, standards, or a regulation. Regulatory inspections -- conducted by an independent state-run body -- are external to the deploying organization. Conformity assessments can be internal or external. In contrast to third-party audits, conformity assessments aim to achieve compliance, while third-party audits by researchers and civil society aim to prevent potential harm or uncover it by systematically evaluating the systems and checking the documents. 

The vast number of actors and contexts in which AI systems function produces multiple complexities for their auditing, and this has affected the precision of related legal provisions. This applies especially when we try to answer questions such as, “What are we auditing for? Which harms count, and how are they defined?”  \cite{institute_algorithmic_2023}. This situation shapes the information we need to assess the system's compatibility with a legal regime or ethical parameter, considering the contextual character of transparency and access to information \cite{asghari_what_2022}. 

We focus on third-party audits, especially those by researchers, academics, and journalists, as well as broader sociotechnical audits.\footnote{These broader types of audits are also referred to as ``ecosystem audits'' by \citet{birhane_ai_2024}.} Such audits are characterized by being conducted post-deployment through systematic evaluations -- mostly a combination of different qualitative and quantitative methods -- and reverse engineering of system behaviour, typically with little data and model access and independent of the organization deploying the algorithm \cite{HCI-083,costanza-chock_who_2022, birhane_ai_2024}. Thus, they have no formal contractual relationship with the audit target \cite{noauthor_mozilla_nodate}.  

In recent years, there has been increasing public and academic concern about the influence of social media on public discourse, democratic processes, and institutions \cite{Bail.2018,quattrociocchi_social_2017}. Over the years, researchers have encountered significant challenges when attempting to access online platforms' data for the purpose of studying their operations, potential risks, and impacts \cite{olteanu2019social,ohme_digital_2023}. Moreover, access to AI technologies beyond ML-driven social media recommendation systems, such as ADMs or general-purpose AI, presents a significant challenge \cite{institute_algorithmic_2023}. Scientists and civil society are dependent on the voluntary granting of access, and sometimes only process access is granted, i.e. access to documentation (e.g \citet{allhutter2020ams}). Are we repeating the same mistake by granting large companies unwarranted authority without enabling civil society to thoroughly examine and understand their products for potential harm?

In response to these challenges, the DSA and the AIA have introduced distinct frameworks for audits, third-party audits and access to these technologies by research and civil society. In the following section, we will introduce and explore both of these frameworks, shedding light on how they address (third-party) audits of platforms and AI products.
\section{The Regulatory Framework Based on the DSA and AIA on Third-Party Audits and Data Access}
\label{sec:framework}
The objective of this chapter is to comprehend the regulations of third-party audits and third-party data and model access by the DSA and AIA. We will begin by examining the DSA's provisions on audits and data access (section \ref{sec:dsa}). Then, we will do the same for the AIA based on the \textit{Corrigendum} made public by the European Parliament (section \ref{sec:AIA}).
\subsection{Digital Services Act}
\label{sec:dsa}
The DSA aims to create a harmonized legal framework that is applicable to all online intermediary services provided in the EU. The regulation helps promote a safer digital space by regulating the distribution of (illegal) content. To increase accountability to the public, the regulation includes a number of specific obligations for providers of VLOPs and VLOSEs \cite{european_commission_dsa_nodate}, including provisions such as specific transparency obligations on content moderation or the creation and publication of reports on risk assessments and mitigation procedures. 
\subsubsection{The Regulations on Independent Auditing in the DSA, Art. 37 DSA}
In order to ensure compliance with the DSA, a “novel institutional ecosystem” \cite{laux_taming_2021} will be established that involves independent third parties in an additional oversight role.\footnote{Cf. Recital No. 92 DSA.} \cite{dsa} Therefore, the officially published version of the DSA asserts that VLOPs and VLOSEs are legally required to undergo an independent audit at least once a year (“independent audit”)\footnote{Art. 37 (1) DSA.} The purpose of auditing is to determine whether the providers are following the obligations standardized by the DSA.\footnote{It means the obligations set out in Chapter III (Art. 11-48 DSA), in particular the obligations arising from codes of conduct referred to in Art. 45 and 46 DSA and the crisis protocols referred to in Art. 48 DSA.} This is an important regulatory requirement, as it increases providers’ accountability and enables regulators and the public to understand and regulate how VLOPs and VLOSEs moderate content. 
The Commission is empowered to implement the requirements of Art. 37 DSA by legislative act. It has published a first draft of a regulation to specify the requirements for independent audits (hereinafter referred to as the Delegated Regulation on independent audits) \cite{euregulation}. The introduction of such guidelines is welcome as it will provide legal certainty for both audit firms and VLOPs and VLOSEs by providing guidance that can improve transparency and accountability.
An important audit tool is the preparation of an audit report, which should comply with a comprehensive set of obligations. The audit firm must describe the specific elements audited and the methodology used.\footnote{Art. 37 (4)(2d) DSA.} It must also provide a description and summary of the main audit findings.\footnote{Art. 37 (4)(2e) DSA.}  In addition, an outcome statement must be given as to whether the audited provider has complied with the obligations and commitments referred to above.\footnote{Art. 37 (4)(2g) DSA.}  The audit statement shall state the result, which can be “positive” \footnote{“Positive" means, where the auditing organisation concludes with a reasonable level of assurance that the audited provider has complied with an audited obligation or commitment, Art. 8 (1)(a) Delegated Regulation on Independent Audits \cite{euregulation}.}, “positive with comments”\footnote{“Positive with comments" means, where the auditing organisation concludes with a reasonable level of assurance that the audited provider has complied with an audited obligation or commitment, Art. 8 (1)(b) Delegated Regulation on Independent Audits.} or “negative”\footnote{“Negative" means, where the auditing organisation concludes with a reasonable level of assurance that the audited provider has not complied with an audited obligation or commitment, Art. 8 (1)(c) Delegated Regulation on Independent Audits.} In the case of a negative remark, the audit firm must make operational recommendations to be implemented within a recommended period of time to achieve compliance.\footnote{Art. 37 (4)(2h) DSA} If the audit company cannot conduct the auditing at all or can only do so partially, this should be clarified.\footnote{Art. 37 (5) DSA.} In such cases the Delegated Regulation on independent audits could specify how the systematic assessment of risks must be conducted. It should also specify the factors to be taken into account in the risk analysis.\footnote{Cf. Art. 9 (4) Delegated Regulation on independent audits.} According to the current draft, the inherent risks\footnote{“Inherent risks" means the risk of non-compliance intrinsically related to the nature, the activity and the use of the audited service, as well as the context in which it operates, and the risk of non-compliance related to the nature of the audited obligation or commitment, Art. 2 (10) Delegated Regulation on Independent Audits.}, the control risk\footnote{“Control risks" means the risk that a misstatement is not prevented, detected and corrected in a timely manner by means of the audited provider’s internal controls, Art. 2 (11) Delegated Regulation on Independent Audits},  and the detection risks\footnote{“Detection risk" means the threshold beyond which deviations or misstatements by the audited provider, individually or aggregated, would reasonably affect the audit findings, conclusions and statements, Art. 2 (12) Delegated Regulation on Independent Audits.} should be accounted for in the audit companies’ analysis.\footnote{Art. 9 (3) Delegated Regulation on Independent Audits.} 
Providers should provide the necessary cooperation and assistance to organizations conducting audits under this obligation so that they can undertake their work in an effective, efficient, and timely manner.\footnote{Art. 37 (2)(1) DSA.} They must, for example, grant access to all relevant data and premises. This includes the disclosure of data relating to algorithmic systems and answering written or oral questions.\footnote{Cf. Recital No. 92 DSA. In addition to this, the Delegated Regulation on independent audits could regulate which information the providers have to provide to the audit organisations, Art. 5 (1) Delegated Regulation on Independent Audits. The provider and the audit organisation must necessarily agree on conditions - i.e. meaning duties and obligations - that are listed in the current draft to carry out the audit, Art. 7 (1)(a) Delegated Regulation on Independent Audits.} The platform provider must then consider improvement suggestions and take action to enforce the recommendations within one month.\footnote{Art. 37 (6)(2) DSA.} If the operational recommendations are not implemented, the VLOP or VLOSE provider must justify and specify alternative measures that will be taken to fix the identified cases of non-compliance.\footnote{Art. 37 (6)(3) DSA.} 

\subsubsection{The Requirements for Audit Firms under the DSA}
The auditors should have no conflicts of interest with the provider.\footnote{Art. 37 (3a) DSA. In this respect, the European legislator clarifies that audit firms are organizationally independent from the interests of the provider and constitute an external audit structure.
The additions were explicitly included to address concerns that audit firms are dependent on platform providers for auditing. Cf. critical comments on the preliminary draft \cite{buri-van-hoboken-dsa-observatory}.} Furthermore, they must have proven expertise in risk management as well as technical competence and capability. \footnote{Art. 37 (3b) DSA. This requires that they must have the necessary knowledge in the field of risk management as well as the technical expertise to test algorithms, cf. Recital No. 92 DSA.} Providers are allowed to choose the audit firms themselves.  However, it is important to note that the DSA introduced a rotation model according to which the audit firms may only work for the providers if they have not performed an audit pursuant to Art. 37 DSA for a period of more than ten consecutive years.\footnote{Cf. Art. 37 (3a)(ii) DSA.} 

\subsubsection{Data Access for Vetted Researchers, Art. 40 DSA}
Art. 40 DSA regulates access to the data of VLOPs and VLOSEs for ``vetted researchers.'' The norm contains not only the obligation of providers to grant access, but also the right of researchers to access. \footnote{E. g. \cite{kaesling-dsa} and \cite{wehde-dsa-e}} Researchers must make their research results publicly available free of charge \footnote{Art. 40 (8)(1g) DSA.}. The results can thus be made available to the authorities, the Commission and the public. Classification as an approved researcher is subject to strict conditions. In particular, researchers must be affiliated to a research institution \footnote{Art. 40 (8)(1a) DSA.}. It means a university, including its libraries, a research institute or any other entity, the primary goal of which is to conduct scientific research or to carry out educational activities involving also the conduct of scientific research. \footnote{Art. 2 (1) Directive (EU) 2019/790.} Like the audit firms, the researchers and the institutions to which they belong must be independent of commercial interests \footnote{Art. 40 (8)(1b) DSA.}. Another condition for access is that the data collection is necessary and proportionate for the creation of the data work. \footnote{Art. 40 (8)(1e) DSA. Cf. Recital 97.} The classification as vetted researcher is granted to the researcher only in relation to the intended research project.  The research must relate to systemic risks and contribute to their identification and understanding.\footnote{Art. 40 (4) and (8f) DSA.} 

\subsection{The Artificial Intelligence Act}
\label{sec:AIA}
The AIA \cite{AIA,EU-AI-Report} aims to regulate and harmonise the marketing, commissioning and use of AI systems within the European Union. To this end, safe and trustworthy AI is to be created by observing the fundamental values protected in the European Union, that is material and ethical values, during development and use. For this reason, companies\footnote{The regulation is addressed to all companies that place AI systems on the market or put them into operation in the European Union, cf. Art. 2 (1) AIA \cite{AIA,EU-AI-Report}. In practice, this will include software developers, AI importers and distributors, but also, for example, government authorities.}  should be subject to a far-reaching catalogue of obligations and ensure quality assurance throughout the entire AI value chain.
\subsubsection{The Regulations on the Conformity Assessment Procedure}
The AIA only regulates conformity assessment requirements for high-risk AI systems. For this purpose, the AIA defines ``conformity assessment'' as ``the process of demonstrating\textit{\textbf{ }} whether the requirements [...] to a high-risk\textit{\textbf{ }}AI system have been fulfilled.''\footnote{Art. 3 (20) AIA.} The aim is thus to promote the accountability of AI providers. On the one hand, a so-called ``embedded'' AI system\footnote{Art. 6 (1) AIA and Annex I. These are, for example, products for machines, medical devices, automobiles and aircraft, Rec. (50) AIA.} is subject to conformity assessment by third parties. AI systems are already subject to other product safety regulations; the conformity assessments are carried out in accordance with these different European regulations. \footnote{Art. 43 (3)(1) AIA. Cf. Rec. (49),(51) AIA.} In this case, however, the provider of the AI system must take into account some of the procedural requirements of the AIA.\footnote{Art. 43 (3) AIA.} As these AI systems are already subject to detailed safety requirements, which are assessed under a public authorisation procedure, the AIA is intended to complement these requirements.\footnote{The regulation of the systems is based on the legal provisions of the New Legislative Framework. This means that the product safety requirements for a system should be based on an EU Directive, with the standardisation organisations CEN, CENELEC and ETSI providing more specific details.} For so-called stand-alone AI systems\footnote{Art. 6 (2) AIA and Annex III.}, the conformity assessment must generally be carried out internally, i.e. by the providers themselves. With regard to these systems, it is explicitly stated that the involvement of a so-called notified body is not necessary for the assessment.\footnote{Art. 43 (2) AIA.} Due to the complexity of these AI systems and the involved risks, the legislator assumes that the suppliers are best placed to evaluate them. This also means that audits that potentially are part of such a conformity assessment can be conducted internally, thus implying first-party auditing. However, this does not apply to AI systems that are used as "remote biometric identification systems" (Annex III No. 1(a) AIA.). For these, it is optional whether the conformity assessment is carried out by the provider himself or by a notified body.\footnote{Art. 43 (1) AIA.} Furthermore, the conformity assessment by external bodies is mandatory for a few cases, e.g. when the AI system is put into operation by law enforcement or immigration authorities or by institutions, bodies, offices or agencies of the Union.\footnote{Cf. Art. 43(1), (2) AIA. For general purpose AI systems, cf. Rec. (161) AIA.} Another requirement for specific 'high-risk' AI systems \footnote{Cf. Art 27 (1) AIA. Mandatory for high-risk AI system as described in Article 6(2), with the exception of those intended for use in the areas listed in point 2 of Annex III, deployers that are bodies governed by public law, or are private entities providing public services, and deployers of high-risk AI systems referred to in points 5 (b) and (c) of Annex III.} is the execution of a fundamental rights impact assessment, however, also conducted by the deployers similar and an extension to data protection impact assessments \cite{thelisson_conformity_2024}. 
\subsubsection{The Requirements for Audit Firms under the AIA} 
A 'notified body' is defined in the AIA as ``a conformity assessment body notified in accordance with this Regulation and other relevant Union harmonisation legislation.''\footnote{Art. 3 (1)(22) AIA. For the term 'conformity assessment body', see Art. 3 (1)(21). Cf. for the term 'notifying authority' Art. 3 (19) and Art. 28 (1) AIA. For the duties of the notifying authority, cf. Art. 3 (19) and 28-30 AIA.} The procedure for their designation sets out  requirements for their (professional) qualifications as well as their independence and objectivity. As in the DSA, the AIA also clarifies that notified bodies should be independent of the provider of the high-risk system. This includes both economic aspects, e.g. no competitive relationship between the provider of the AI system and the notified body, and personnel aspects, e.g. no connection between the management and the AI development project.\footnote{Art. 31(4) and (5) AIA.} In order to ensure technical competence, they have to note  specific organisational, quality management, personnel and procedural requirements.\footnote{Art. 31(2), (3), (10) and Rec. (126) AIA.} To this end, notified bodies should have procedures in place to guide their activities, taking into account the size of an undertaking, the sector in which it operates its structure, and the degree of complexity of the AI system concerned.\footnote{Cf. 31(8) AIA.} In addition, it should be noted that the conformity assessment procedure for AI systems, that will be used by certain authorities and related bodies (e.g. law enforcement, immigration or asylum authorities or EU bodies) is carried out by the market surveillance authority.\footnote{Cf. Art. 43(1) AIA.} 
Unlike the DSA, the AIA ensures the cooperation between the developer, the notified body, and the notifying authority. The authority is enforced to certify the bodies. In addition, the certificated bodies are accountable to the notifying authorities. To qualify as a notified body under the AIA, the potential audit firm must fulfill several requirements: The notification application must be accompanied by comprehensive information, such as the conformity assessment activities, conformity assessment modules, and AI technologies for which the conformity assessment body claims competence, as well as - if available - an accreditation certificate.\footnote{Cf. Art. 29 (2)(3) AIA.}
The provider shall ensure that the conformity assessment process takes place.\footnote{Cf. Art. 6(f) AIA.} An audit should also be carried out if the AI system has already undergone a conformity assessment procedure, but a significant change has been made to the system.\footnote{Cf. Art. 43(4) AIA, for the definition in Art. 3(15) AIA.} In any case, they are obliged to document the assessment in order to provide evidence that the conformity assessment has been carried out properly.\footnote{Art. 18 (1)(g), Art. 22 (3)(b) AIA.} The requirements for the conformity assessments by internal controls are set out in Annex VI. It consists of the establishment and monitoring of the relevant quality management system,\footnote{Annex VI No. 2, No. 4 AIA} as well as the control of the obligations imposed on the suppliers regarding the technical documentation.\footnote{Annex VI No. 3 AIA}
The providers of the AI system should assist the testing organisations in their investigations, e.g. by making available the technical documentation that provides all the information necessary to assess whether the AI system meets these requirements.\footnote{Cf. Rec. (86) AIA.} Providers must give them full access to the training and test data sets used. This includes application programming interfaces and other means and tools suitable for remote access.\footnote{Annex VII No. 4.3 AIA.} For example, the audit firm may, upon request, be given access to the AI system's training models and trained models, including the relevant parameters, and should be able to test the AI system independently.\footnote{Annex VII No. 4.4 and 4.5 AIA.} For this reason, it is appropriate for the notified body to maintain the confidentiality of the information provided.\footnote{Cf. Art. 31(7) and Art. 78 AIA.} If the notified body can confirm the conformity of the high-risk system, a certificate - similar to the audit statement under the DSA - is issued.\footnote{Art. 44(1) and (2) AIA.} This certificate is valid for up to five years or is revoked or restricted if the AI system subsequently fails to comply with the requirements of the AIA.\footnote{Art. 44(3) AIA} In order to adapt the legal framework, the Commission is empowered to adopt instruments amending the provisions on conformity assessment procedures.\footnote{Art. 43(6) and Art. 97, Rec. (173) AIA.}
\subsubsection{No Data Access for Vetted Researchers}
The AIA does not contain a regulation comparable to Art. 40 DSA. This is surprising, as the regulation aims to ensure that relevant entities, such as digital innovation hubs, testing experimentation facilities and researchers, should be able to access and use high quality datasets within their respective fields of activities for the development and assessment of high-risk AI systems.\footnote{Rec. (68) AIA.} The European data space created by the AIA and the facilitated exchange of data should also provide trustful, accountable and non-discriminatory access to high quality data for the training, validation and testing of AI systems.\footnote{Recital No. 45 AIA.} 
\section{Creating a Diverse AI Audit Ecosystem for Accountability by Involving Research and Civil Society}
\label{sec:ecosystem}

In this section, we describe an AI audit ecosystem that ensures two key accountability components. For that, we draw on a definition of accountability in the AI context. We look at three aspects which demonstrate why internal audits are not sufficient to create an audit ecosystem. In three additional aspects, we point out why third-party audits are valuable for an AI audit ecosystem . This will be the blueprint under which we will explore whether the analyzed EU legislation can create an AI audit ecosystem.

\subsection{Third-Party Audits by Researchers and Civil Society}
\label{sec:third-party}
 
\citet{buolamwini_gender_nodate, ribeiro20202,bloomberg,mattu_how_nodate,sweeney_discrimination_2013} are some examples of algorithm audits in the United States that have uncovered problematic algorithmic behavior or algorithm impacts that could potentially cause harm. Regarding algorithmic audits in the EU, audits of the Austrian AMS algorithm \cite{allhutter2020ams}, and Rotterdam's welfare fraud algorithm by the Wire \cite{wire}, come to mind. These audits were third-party audits, that is, audits by independent, external parties combining independent researchers, civil society, regulators, journalists, and other stakeholders. They were responsible for uncovering deeply-rooted sociotechnical harms in algorithmic systems related mainly to representational harms due to discriminatory design choices. As a result, they showed the extent of the potential harms brought by AI systems' biases.

With respect to social media platforms, third-party audits uncovered social system harms due to the propagation of disinformation on social media, hate speech increase on Twitter \cite{hickey2023auditing} and uncovered bias with respect to right-winged content recommendation on YouTube \cite{ribeiro20202}. It has to be noted that third-party audits by researchers and civil society are not conventionally considered audits per se due to their narrow target specification \cite{raji_outsider_2022}, but they have repeatedly shown their significance. Our argument, therefore, favors their integration in an AI audit ecosystem. 

Third-party audits aim for a material change in the situation (e.g., a product update, policy change, or recall) to minimize the harm experienced by those they represent \cite{noauthor_mozilla_nodate}. In contrast, first- and second-party audits that seek to validate procedural expectations, aim to minimize liability and solely test for compliance to AI principles and legal constraints \cite{noauthor_mozilla_nodate}.
Despite the important oversight of third-party audits in platform research and algorithmic systems, it has been difficult for them to audit these models formally and legally. Until now, evaluations are usually done through the application developers’ voluntary public disclosures of possibly incomplete transparency reports or by voluntarily providing certain data access through an API \cite{albert_john_diverse_2023}.

\subsubsection{Third-Party Platform Data Access}
\citet{olteanu2019social} reports four obstacles with systematic platform research through third parties: (1) Many social platforms discourage data collection by third parties as some platforms such as Facebook block their API for researchers and other platforms that block scrapers.  Additionally, there is a shift toward APIs of major platforms and products being closed, a scenario described as the ``post-API scenario'' \cite{trezza_scrape_2023}. This leads to more open-source products or products being audited by third parties by deployers, who are probably more concerned about the compliance of their products. This leads ultimately to an imbalance in oversight \cite{raji_outsider_2022}. (2) Even
API access comes with limitations as they limit the quantity of data, increase the monthly costs and provide only query languages with limited expressiveness. 
(3) Additionally, these platforms may not give access to all data as they safeguard privacy-related content although the platforms could anonymize them. 
(4) Finally, the platform may revoke access every time. 

For example, access for researchers has changed recently for the Twitter API \cite{twitterapi} and the Reddit API, as both have raised the costs for third-party API access \cite{redditapi}. Nevertheless, attempts by third parties such as NGOs, research and investigative journalists to expose these accountability concerns by data donations or scraping are thwarted by companies. For example, the NGO AlgorithmWatch was prevented to conduct an audit of the Instagram algorithm via data donations. Facebook sued them for disregarding the terms of service \cite{instagramalgorithmwatch}. Data donations are only necessary because the APIs for limited 'black-box' access have been shut down or made considerably more expensive by many platforms and services \cite{redditapi,twitterapi,trezza_scrape_2023}. 

The DSA will provide data access for vetted researchers. However, as demonstrated vetted researchers may not include NGOs and investigative journalists, although many of the case studies mentioned were conducted by newspapers and were accompanied by NGOs that were able to contribute with their domain knowledge to potential harms. We will discuss reasons for this being critical later.

\subsubsection{Third-Party ADM and ML-Model Access}
It appears to be just as challenging with ADMs and ML-driven products. Many of the products do not have access for researchers and regulatory audits. Therefore, mostly commercial APIs have to be used as in the Gender Shades case \cite{buolamwini_gender_nodate}. If there is no API, the data can only be scraped, which is mostly illegal or prevented by other means. US Researchers are currently campaigning against the use of anti-hacking laws in the US that prevent audit studies by scraping. At the same time, according to \citet{raji_outsider_2022}, companies continue to seek to prevent audits by (1) paywalls, (2) prohibition laws through terms of services, and (3) structuring the product to obscure any clear set of test points. \citet{raji_outsider_2022} describes it as an extraordinary effort for researchers, NGOs, investigative journalists to gain access and knowledge of ML and ADM systems. 

It was shown that the AIA does not encompass data or model access provisions for third parties, even in the context of high-risk systems. Nonetheless, in the subsequent discussion, we will explore the potential access levels and, consequently, the knowledge accessible to third parties concerning AI technologies. Based on these findings, we will determine to what extent it is possible to make statements about a system even with limited access.

\subsubsection{Technical Methods and Access Levels}
\label{sec:access}
Third-party audit methods \cite{sandvig_auditing_2014, bandy_problematic_2021, the_ada_lovelace_institute_technical_2021,mesmer_auditing_2023} -- such as code audits, user audits, automated audits (including scraping audits, sock puppet audits and API audits), document audits, crowdsourced audits, and experimental audits \cite{the_ada_lovelace_institute_technical_2021} -- contribute various approaches to auditing platforms and AI systems. Whether social media platform, ADM or ML model audits, the methods for third-party audits are similar.

Document audits are often broad sets of information that can be analysed in different ways \cite{mesmer_auditing_2023} and ensure the possiblity for third-party impact evaluations or conformity assessments as intended by the AIA. However, only process-based compliance is feasible through document audits, there are no standards for most of these reports and they are impossible to verify if no other data access is provided  \cite{mesmer_auditing_2023}.

Code audits entail direct access to a system’s code and involve the dynamic execution of code with various inputs to observe corresponding output behaviors \cite{the_ada_lovelace_institute_technical_2021}. While they are suitable for some audits, code audits might not be ideal for AI product audits because of the need for training data. Furthermore, potential complexities in reverse engineering system behavior - especially for dynamically trained algorithms such as recommender systems - hamper code audits' feasibility. 

Scraping audits, on the other hand, lack randomization and interventions. This limits their utility primarily to descriptive tasks \cite{bandy_problematic_2021}. They additionally lack personalization for specific user accounts \cite{mesmer_auditing_2023}. Furthermore, scraping and sock puppet audits often violate terms of services of platforms \cite{mesmer_auditing_2023}. API access limitations were already discussed. 

Experimental audits involve the change of the algorithm and can thereby establish causal evidence since interventions are feasible \cite{gaddis_introduction_2017}. However, full access to the system is necessary for experimental audits, and experiments with social platforms such as Facebook on emotional contagion exhibit ethical difficulties \cite{selinger_facebooks_2016}. 

Crowdsourced audits, including data donations, have two major advantages: They include users in the audit, and they transfer parts of the training data so that more statements are possible. However, in the past, large platforms or companies have prevented such crowdsourcing audits as in the case of AlgorithmWatch.  Recognizing the inherent advantages and limitations across various audit types, a comprehensive analysis of these factors would extend beyond the scope of this article\footnote{For a comprehensive analysis of connections between audit types and recommender system elements see \cite{mesmer_auditing_2023}.}. Consequently, our focus is confined to examining access to AI systems.

Technical access and the knowledge of a system can be categorized on a continuum from ``white-box'' to ``black-box'' and is related to these audit types. \citet{koshiyama_towards_2021} have defined seven levels of auditing access that range from “process-access,” where only indirect observation of a system is possible, and the system’s behavior in a sociotechnical system must be reverse-engineered, to “white-box,” where details of the encompassing model are disclosed. The latter is usually not applicable to third-party audits, as it would conceal all information about the model, which could have security, privacy, and trade secret issues \cite{koshiyama_towards_2021}. Thus, \citet{koshiyama_towards_2021} have described an “information concealed versus feedback detail trade-off.”

\begin{table}[]

    \centering
     \resizebox{\columnwidth}{!}{
    \begin{tabular}{p{1.5cm}||p{3cm}|p{4cm}|p{3cm}|p{3cm}|}
    \cline{2-5}
    & What is accessed & Tools and possible evaluations&Feasible audit types&Case study third-party audit(s)\\
   \cline{2-5} \cline{2-5}
   \rowcolor{Gray}
    Process access & Frameworks, checklists, model Cards, datasheets, method paper, technical reports&Checklists for explainability, robustness, fairness and privacy& Document audits (algorithmic risk
assessment and algorithmic impact evaluations) &Audit Austrian AMS algorithm\cite{allhutter2020ams}\\
    
    Model access ('black-box')& Access to the predictor model $f(\cdot)$ using artificial data $x^{\ast}$ &Feature relevance plots, partial
dependency plots, adversarial
attacks, adversarial
fairness, statistical disclosure& Scraping, API, sock puppet audit and code Audit&Bloomberg stable diffusion audit\cite{bloomberg}, discrimination in online ad delivery, \cite{sweeney_discrimination_2013}, auditing radicalization pathways on YouTube \cite{ribeiro20202}\\
 \rowcolor{Gray}
    Input access & Access to the predictor model $f(\cdot)$ and training data $x$ to predict $f(x)$&Surrogate
explanations, synthetic data, bias in
outcome&Crowdsourced Audit\footnote{just a fraction of training data $ x_p \in x$}, API audit with training data &Cooperative audit of pymetrics' candidate screening software\cite{10.1145/3442188.3445928}\\
   
    Outcome access ('grey-box')& Access to the predictor model $f(\cdot)$, training data $x$ to predict $f(x) = \hat{y}$ and the actual outcome $y$&Accuracy of
explanations, concept drift
analysis, bias in
opportunity, inversion
attacks&API audit with training data and history of actual outcomes&ProRepublicas investigation of North-
pointe’s recidivism risk scoring system \cite{mattu_how_nodate}\\
     \rowcolor{Gray}
    Parameter control and learning goal access & Learning procedure and target $L$, parameters $\phi$, $L(f_\phi(x))$ can be retrained& Stability of
explanations, stability analysis, stability of bias
metrics, functionality
stealing, model
complexity, stress-testing, trade-off of
bias and loss
metric, model
extraction& Retraining of the actual model with different parameters &Audit of the Rotterdams' welfare fraud ADM\cite{wire,noauthor_suspicion_nodate}\\

    White-box access& Access to the whole architecture $f$ and all documents &Documents
and specific
explanations, model
selection and
validation, model
selection and
development, model
security
evaluation&Experimental audits (interventions) &  Accessible by first-parties and possibly second-parties\\\hline
    \end{tabular}
    }
    \caption{The Table presents leveled access modes from 'process access' to 'white-box access'. We connected access modes with 'what is accessed', tools and feasible audit types. The table must be read so that in each access mode possible tools, audit types and accesses are added to those that were in the level before. This may not be the case only for process access, since documents are not available in each access level.
    The table is inspired and derived from \citet{koshiyama_towards_2021}, \citet{9755237} and \citet{brown}. As in the case of table \ref{tab:characteristica}, this table should provide only an overview.  We derived the table via literature review and not via systematic methods. A systematic evaluation may not be feasible as the field is still emerging. However, we aim to illustrate the connection between data access and audit methods and demonstrate the trade-off between access and revealed information.}
    \label{tab:access}
\end{table}

Table \ref{tab:access} shows different access levels that \citet{koshiyama_towards_2021} have proposed for ML algorithm audits. We matched them with third-party evaluation tools, feasible audit types, and case study third-party audits. This representation is a simplification because platforms and general-purpose AI consist of several models that interact with each other and ADMs can consist of several models and a complex decision function. Nevertheless, the presentation is useful to show three aspects. First, with more access, more knowledge about a system is possible and more audit types are available. Second, there is a trade-off between trade secrets and model access. Third,  case study third-party audits in the past that had only 'black-box' access -  mostly through an API, scraping, or sock puppets - demonstrate that low-level access can contribute to accountability.

Some audits, such as the Austrian AMS algorithm relied solely on process access. The prediction model has not been made public \cite{allhutter2020ams}. Generally, releasing personal data relating to ADMs would obviously be dangerous. In this case, anonymization methods would have to be used to release the data for scientific purposes like it is planned by the DSA. These document audits have little expressiveness as they have to reverse engineer the whole system's functioning without the knowledge or verification that the documents are actually accurate. Others with at least 'black-box' access could provide more information. 

For example, Bloomberg's stable diffusion audit \cite{bloomberg}, or in YouTube's radicalization audit \cite{ribeiro20202}, the prediction model was accessed via the API. Thus, problematic behaviour like biases could already be shown in both cases. The original Gender Shades study as well operated within the constraints of accessing their audit targets via a commercial API, mirroring the actions of a user within these platforms and thus using a socket puppet \cite{raji_actionable_2019}. Others had even more access. In the case of the cooperative audit of pymetrics' candidate screening software the auditors had `grey-box' access, as the outcome is not feasible\footnote{As it's unknowable if someone would have performed well that did not get the job.} for recruitment algorithms\footnote{This audit was criticized as \cite{10.1145/3531146.3533194} claimed that the authors got funded by the company for reducing the scope of the audit. If this claim is justified, this would mean that the audit would not been conducted by independent external auditors, but dependent and thus internal auditors. The example again demonstrates the power of tech companies.}\cite{wilson_building_2021}. In the case of ProRepublicas investigation of the recidivism risk scoring system they had access to the model, the training data and the actual outcome \cite{mattu_how_nodate}. In the case of  Rotterdam's welfare fraud audit, the auditors had parameter control and learning goal access \cite{wire}. Model access alone has shown its effectiveness through certain case study audits. Audits, including Bloomberg's stable diffusion audit \cite{bloomberg}, examinations of discrimination in online ad delivery  \cite{sweeney_discrimination_2013}, and the audit of radicalization pathways on YouTube \cite{ribeiro20202}, have achieved auditing outcomes with limited access, albeit frequently encountering challenges due to legal constraints. Especially in the case of Gender Shades, where after the audit there was an improvement in several facial recognition tools in terms of functionality for black women \cite{raji_actionable_2019}. This proofs limited 'black-box' access for researchers can contribute to mitigating harm. 

While other levels of access might provide more comprehensive insights into a system, the prevalence of reverse-engineering methods underscores the significant influence of third-party audits. Admittedly, a system-tailored approach is necessary, given the nuances involved, and model-agnostic audits, such as partial dependency plots and feature relevance plots, present challenges \cite{hansen_model-agnostic_nodate}. Nonetheless, it is evident that third-party audits, even with 'black-box' access, have managed to provide insights when combined with evaluations of the sociotechnical framework. Particularly because they contribute to a system for which the depth of knowledge acquired corresponds to the access level, audits with even minimal access can generate momentum for obtaining greater access. In some cases, this may involve collaborating with other auditors and stakeholders. This concept is further explored in the next section (section \ref{sec:aiaudit}) It is worth noting that substantial examination by sock puppets or API audits is often feasible only for products that possess open APIs.


\subsection{An AI Audit Ecosystem}
\label{sec:aiaudit}
Algorithmic accountability regards a networked account for an algorithmic system, where several actors must explain and justify to a forum or several fora their use or design of an algorithmic system, as well as the decisions and the subsequent effects concerning the system. The fora can be internal and external, as well as formal and informal \cite{wieringa_what_2020}. \citet{NovelliForthcoming-NOVAIA} have defined the goals of AI accountability as (1) compliance, (2) reporting, (3) oversight, and (4) enforcement. 
Although audits interfere with reporting and enforcement, they primarily achieve compliance and oversight. Compliance concerns the binding of the system to align with ethical, legal, or technical norms. For example, performing an ethical assessment may constitute compliance. Typically, compliance refers to compliance with the law. Oversight seeks to find relevant facts about the system, such as performance, bias metrics, or other measures, through an overseeing body that can be internal or external.

In order to ensure oversight and compliance, \citet{raji_outsider_2022} and \citet{albert_john_diverse_2023} have proposed establishing an AI audit ecosystem that includes internal and external auditors. \citet{sambasivan_non-portability_2020} have called for the establishment of an ecosystem for accountability in India that includes various stakeholders and empowers oppressed communities. \citet{mokander_conformity_2022}, meanwhile, have argued that the AIA de facto proposes to establish an AI audit ecosystem. However, none of these authors have described what constitutes an audit ecosystem or why we need one. While following this line of reasoning about the importance of including diverse auditors, we define an audit ecosystem and argue that the AIA does not establish an AI audit ecosystem. 

We define an AI audit ecosystem as an open dynamic partial self-organizing community of hierarchically independent, yet interdependent heterogeneous auditors and audit types\footnote{Inspired by ecological, digital and innovation ecosystems: "a community of hierarchically independent, yet interdependent heterogeneous participants who collectively generate an ecosystem output"\cite{thomas_innovation_2019}.}. The characteristics are described by the following: 

\begin{itemize}
    \item open, meaning that it interferes with other accountability processes, such as reporting, enforcement, and harm incident reporting 
    \item dynamic over time (and thereby adaptive to new forms of AI systems, which may have to be audited differently) and over stakeholders (it must involve potentially affected communities and users)
    \item partial self-organization of auditors in an enabling environment\footnote{See in anti-corruption social accountability ecosystems \cite{halloran2021accountability}} and a promotion of sharing audit information through interaction between auditors (formal and informal)
\end{itemize}   

The audit ecosystem is needed to ensure compliance and oversight throughout the AI lifecycle. We argue that for such an ecosystem, third-party audits by researchers and civil society are a vital component to fulfill the characteristics of the ecosystem and thus also provide both oversight and compliance through first-party audits. \citet{NovelliForthcoming-NOVAIA} have stated that ``[o]verseers may act at different levels that may overlap, e.g., an internal audit is compatible with judicial review accountability.'' However, we argue that overseers \textit{should} act at different levels, as (1) internal audits are prone to audit-washing as well as false assurances, and they may only ensure compliance (see section \ref{sec:washing}) and are (2) not adaptive to new technologies (see section \ref{sec:adaptive}). Additionally we argue that (3) NGOs and researchers possess valuable experience in incorporating a holistic perspective that encompasses socioecological-technical systems. They are able to draw from a background in conducting impact assessments and human rights considerations (see section \ref{sec:humanrights}) and (4) tend to include affected communities more than companies as they have other audience and incentives (see section \ref{sec:affeccommunities}).



\subsection{Why Are Internal Audits Insufficient?}


Internal audits and assessments have emerged as a prevalent focus in the academic discourse. In alignment with this trend several regulations -- the Algorithmic Accountability Act of 2022 in the US, the AIA, and GDPR enforcement -- call for first-party audits, underscoring their outsized focus in academia and governance \cite{raji_outsider_2022}. Internal audits have the advantage that the developers know their system best and have ``white-box'' access \cite{koshiyama_towards_2021}. This provides them the most technical knowledge about the system and they can intervene at any stage of the design process as well as throughout the AI lifecycle. It also puts the burden of responsibility for harm prevention on the shoulders of the developers and deployers \footnote{If both parties are the same company.} and thus on those who establish a business model with it. Nonetheless, as demonstrated by precedent audit domains, the adequacy of internal audits in ensuring comprehensive oversight has been called into question, as \cite{raji_outsider_2022} has highlighted.



\subsubsection{``Audit Washing'' and False Assurances}
\label{sec:washing}
 Furthermore, the overreliance on internal audits could lead to unverifiable assertions that the AI systems has passed ethical and legal standards, leading to even more harm and less oversight \cite{goodman_ai_2022}. There is an ongoing discussion within research communities on how to translate ethics into practice. The discussion evolves around which fairness metrics should be used as fairness remains a contested term \cite{munn_uselessness_2023}. General standards and norms for audits \cite{costanza-chock_who_2022} are therefore absent. This leads to an environment conducive to the phenomenon of ``audit-washing''. A firm's business goals may not always align with harm-reduction through course-correction or ethical design of their products. Developers and deployers may ignore audit recommendations that threaten
their business interests \cite{mokander_auditing_2023}. 

Empirical research substantiate these concerns. \citet{mcnamara_does_2018} have discovered that instructing developers to incorporate ethical codes yielded no notable change in their established practices. Moreover, studies involving AI development companies \cite{vakkuri_ethically_2022} and startup-like environments\cite{vakkuri_this_2020} have shown a disconnect between acknowledging the significance of ethics and integrating them into AI practices, indicating a substantial gap between ethical research and practical implementation and auditing. 

Such absence of standards and norms in corporate environments could lead to using tools and metrics that play down the risks of ones own product \cite{institute_algorithmic_2023}. Statistically, through \textit{p-hacking} or \textit{data dredging}, and more generally through \textit{fairness gerrymandering}, it may always be possible to suggest that your system is fair \cite{bandy_problematic_2021}. Issues concerning independence of second-party auditors were discussed frequently in financial audits. The opinion paper on this issue of \citet{bazerman1997impossibility} highlights the the emerging bias towards the client in contractual relationship audits. In addition to intentional false assurances of the absence of harm, audits may be performed incorrectly or issues may go unaddressed. That is what happened in the Gender Shades audit \cite{buolamwini_gender_nodate}. It revealed first assessment and benchmarks by the U.S. National Institute of Technology (NIST) failed to measure demographic differences in their regular Facial Recognition Vendor's Test  \cite{raji_outsider_2022}. 

Several possible adjustments address for these issues. \citet{costanza-chock_who_2022} have proposed mandating (1) the publishing of (first-party) audit results, (2) harm incident reporting and other measures to detect and report harms, (3) the involvement of affected parties, and (4) the consideration of accreditation and standards for evaluation. Although all points are valuable for an audit ecosystem, we argue that the involvement of affected parties and standards for evaluation are only valuable if third-party audits support the ecosystem. (1) and (2) are particularly important for an open and balanced ecosystem, as information-sharing is promoted, and multiple self-organized auditors can immediately note and review potential harm incidents. While (4) is especially important for functioning enforcement, (3) is a basic condition for a dynamic audit ecosystem; otherwise, possible harm remains unnoticed.

\subsubsection{Standards, Gaming the System and Goodhart's Law}
The development of standards is vital for compliance and verification. 
However, hard standards and technical targets (e.g. bias metrics) can also lead to decoupling from the intended purpose. In financial audits, decoupling was described as an audit process that becomes ''ceremonial" and rote. This can happen when the entity undergoing auditing has internal control structures that ritualize and channel outside audits \cite{Power1999-POWTAS-3}. 

Gaming the system is connected to Goodhart's law, which is usually paraphrased as "When a measure becomes a target, it ceases to be a good measure." Thus, behaviors are created to achieve the metrics while pursuing other business incentives simultaneously. It could encourage adherence to metrics while maintaining business. Methods to game the system emerge to avoid implementing the desired behavior measured by the metrics \cite{Power1999-POWTAS-3}. For example, in the financial crisis, banks went bankrupt, which had been recently testified solvent \cite{Sikka2009financial}. 

The Wirecard scandal in Germany shows that hard standards do not necessarily help to uncover harm. Even though the company was audited several times by different second-party auditors it was still not revealed that the business model of Wirecard was a financial fraud \cite{lohlein-huber-2022}. In both cases, the companies learned to game the system, and game the standards. The Wirecard scandal was ultimately uncovered by investigative journalists \cite{ft-wirecard}. This example demonstrates once again that there should be external oversight by such institutions, which do not have to adhere to hard standards and entrenched audit processes. This is all the more evident when it comes to new technologies that are developing rapidly.
\subsubsection{Non-adaptivity to New AI Technologies Through Hard Standards}
\label{sec:adaptive}
Internal audits and hard standards have several drawbacks. First, social accountability in the public sector has demonstrated that strict process-based evaluations often slow down government action and suppress creativity. Therefore performance-based evaluations in audits are preferable to process-based ones. The role of civil society can be crucial in that regard, as it functions as a vigilant watchdog over both procedural integrity and performance outcomes. The involvement of civil society can introduce innovative performance metrics to supplement existing measures \cite{ackerman2005social}. 

Especially the AI standardization process, as currently envisaged under the AIA, has its challenges with respect to that. In this standardization process, harmonized standards are adopted following a standardization request (mandate) from the Commission to the European Standardization Organizations (CEN, CENELEC, building the CEN-CENELEC JTC21 technical joint committee) for the application of Union harmonization legislation to harmonize AI standards \cite{mcfadden_harmonising_nodate}. 

This process was criticized several times \cite{mcfadden_harmonising_nodate,Worsdorfer2023,laux_three_2024}. First, the process itself is non-participatory in nature \cite{Worsdorfer2023}. Second, the imposition of hard standards on ethical questions that are inherently context-sensitive and dynamic with new technologies is problematic \cite{laux_three_2024}. 

Third, in an audit ecosystem of hard standards -- like in the AIA case -- `gaming the system' can occur from companies trying to exploit those standards and find loopholes. At the same time, new technologies in particular can create standardization gaps. To create an adaptive and dynamic audit ecosystem informal audits without hard standards and processes could help to expose potential harm.

This becomes particularly evident if there is no external oversight from third parties capable of reproducing these evaluations. Novel AI products and technologies could necessitate the development of new evaluation methods. New ethical questions arise, which may have to include ethical trade-offs navigated by local and diverse stakeholder inclusion. For example, in the aforementioned Gender Shades case \cite{buolamwini_gender_nodate} or in recent examples of LLMs and stable diffusion \cite{bloomberg}. Although the relevant legislation and standards do not yet exist, it is difficult to imagine that if evaluation standards had existed, they would have worked for these type of AI systems. Developing standards for evaluating technology will take a long time, which is why it is necessary that third parties with limited access to the model engage in evaluating and understanding emerging systems. This happened with the stable diffusion audit by Bloomberg\cite{bloomberg} \footnote{Reporters used stable diffusion to generate 5,100 images of people using a simple recurring prompt iterating through different categories. Through data-labeling } and with ChatGPT, which researchers extensively evaluated and which showed biases  \cite{ray_chatgpt_2023}.

\subsection{Why Should We Include Researchers and Civil Society in the AI Audit Ecosystem?}

\subsubsection{Third-party Domain Knowledge} 
\label{sec:humanrights}
We emphasized the significance of auditing both the visible external connections and the internal self-regulatory mechanisms of algorithmic systems. NGOs, researchers, journalists, and members of civil society offer domain expertise that is absent in the organizations deploying algorithms \cite{ackerman2005social}. Identifying such contextualities involves a broad range of perspectives about the impacts that such systems might entail. AI experts might not be capable of addressing this issue. 

According to \citet{li}, AI auditors are not always aware of the way digital technologies affect marginalized communities, and their tools for investigating bias, for example, might not necessarily detect behavior that could not be reasonably predicted before its deployment. Instead, the authors write, ``most expert-led auditing methods were developed to detect statistical disparities not, for example, if an algorithm is censuring or harmfully depicting a marginalized community in images or the provision of online services.” This is why we argue that researchers and civil society representatives potentially have specific domain knowledge that include sociotechnical system thinking, as they have knowledge of the human rights impacts that might take place in specific domains exactly for their proximity to marginalized or affected communities. However,  including also the communities impacted by the systems, and not those who research or support them, may also prove fundamental for the advancement of systems that duly protect rights.  

\subsubsection{Inclusion of Affected Communities}
\label{sec:affeccommunities}
Pro-accountability endeavors that encompass diverse interests and ideologies incorporate more legitimacy than those driven by a limited group of professionals \cite{ackerman2005social}. This is why scholars  (e.g. \cite{costanza-chock_who_2022}, \cite{li}) have recommended the inclusion of stakeholders who are or can be affected by AI technologies, or their representatives, in audits. 

We call this a participatory audit\footnote{At this point, it is important to note that the terminology surrounding what we call participatory audits has not yet been consolidated. When affirming the relevance of participatory audits in the context of governments’ accountability,\cite{berthin} refers to “social audits” in a way that is very similar to what we describe as participatory. According to him, they differ from other forms of audits because they are performed by nonexperts and rely on engagement from citizens and/or civil society organizations. It is a mechanism of oversight, of exerting control on government officials to ensure transparency, responsiveness, and effectiveness by putting citizens in the position of active participants, not mere sources of information \cite{berthin}.}, as it consists of a framework that aims to promote the participation of affected stakeholders in different stages of the design, development, and deployment of AI systems. Audits that consider individual or collective accounts of experiences related to algorithmic bias, harm, or injustice through incident reporting are currently predominantly third-party algorithmic audits \cite{rakova_algorithms_2023}. 

Some authors have cited similar approaches for AI auditing as user-engaged approaches; these would consist of the direct participation of users of AI-based applications in surfacing harmful algorithmic behaviors in activities such as those called ``bias bounty'' \cite{globus-harris_algorithmic_2022}. . In this sense, one can frame participatory audits as a set of different methodologies to include users or affected individuals and groups as a whole in AI auditing processes. 

Through methods such as workshops, bias bounty programs, and direct interviews, individuals potentially harmed by these systems can identify problems and help find solutions. That is why civil society organizations, researchers and investigative journalists could benefit from involving affected communities and users in their analysis of these technologies. Besides, AI developers and providers should have the capability to do the same, so as to involve impacted groups from the early design of their systems. The same applies to auditors, who can have their assessments of algorithms from having access to other perspectives in order to identify other impacts posed by these technologies. However, despite the potential benefits, there is lack of incentives for such participatory methodologies to be put into place. That is why regulation and incentives should lead developers, deployers and auditors to take this sort of action. Successful community engagement depends on the willingness of the audited organization to genuinely listen, respond, and act on the concerns and recommendations identified during the audit process \cite{ackerman2005social}. Affected communities may trust third-party auditors more than internal audits conducted by the organization itself. This trust can encourage community members to openly share their perspectives, concerns, and experiences during the audit process.

\subsubsection{Information and Power Asymmetries}
In the past, understanding and evaluating platforms has encountered considerable challenges due to insufficient transparency in their operations despite their vast power and influence. For years, users and regulators were frequently unaware of these digital platforms' internal functioning, inhibiting the effective resolution of misinformation, data protection, and content moderation. Although the DSA points in the right direction by challenging information and power asymmetries, 'audit capture' is possible. The risk is that large tech companies leverage their market power against their new mandatory auditors  \cite{laux_taming_2021}. Because we have let the market power of large tech companies evolve for years without proper regulation. 

A comparable scenario is occurring with AI products with the focus on conformity assessments by the AIA. In general power asymmetries are present in the standardization process, too. As the process of standardizing audits is ongoing, and the extent of public involvement remains uncertain, standards are usually developed by standard-setting organizations in which industry representatives exert significant influence. In contrast, the voices of civil society and consumer advocates are often overheard \cite{castets-renard-ai-act}. Therefore, researchers \cite{j4040043} demand that policymakers take the necessary steps to improve the overall standardization process, including the structural and organizational framework of standardization organizations, to facilitate an inclusive and democratic system that provides for broad stakeholder engagement and dialogue and input on the development of technical standards. Third-party data access is one measure to involve civil society. 

In section \ref{sec:access}, we learned that third-party audits can uncover harms even with limited access. To understand possible harms, researchers must be able to access data. This understanding is crucial for developing new standards. Currently, there is an imbalance where data is collected about both users and non-users of platforms and algorithmic systems, but there is only limited access for researchers to understand how this data is used \cite{VANDEWAERDT2020105436}. Therefore, an audit ecosystem needs to provide leveled access to different auditors who should publish and share their audit results. This leveled access can help reduce these information and power imbalances. 





\subsection{Challenges Of Third-Party Audits}
Third-party audits have some drawbacks. These auditors may lack the technical knowledge or capacity to thoroughly research a system. There is also a risk of revealing secret information or company secrets, which can be prevented by including nondisclosure clauses. However, to protect the auditors, safe harbor clauses should be included as well to prevent them from intentional or unknowingly leaking trade secrets. Companies may fear to share their model and data \cite{costanza-chock_who_2022}. However, there are best-practices in the case of VLOPs via an API and a vetting process. Further, there are techniques available for adequately protecting or anonymizing data while still providing access and other industries like the finance sector where there are legal ptrotections in the case of data leakage \cite{casper_black-box_2024}.

At the same time, one could say that p-hacking \cite{bandy_problematic_2021} allows third parties to select the metrics that indicate harm. However, this is not the same as the first-party audit case, as other researchers could quickly verify the harm. Civil society and social organizations would have no interest in making a bad name for themselves. 

Furthermore, the AI Now Institute argues that third-party audits shift responsibility to researchers and affected communities. However, we cannot rely on these groups to have the resources to audit AI technologies \cite{institute_algorithmic_2023} effectively. We see it as one measure that can work alongside other measures, such as audits through regulators, internal audits, transparency, and the aforementioned accountability processes. 

While solving these problems presents considerable challenges, it is worth noting that the absence of civil society oversight would yield even more unfavorable outcomes. Adequate funding is essential to alleviate the burdens on civil society and research. Without such audits, achieving comprehensive accountability remains inadequate.

Recognizing the difficulties inherent in civil society and researcher audits, it is crucial to acknowledge that algorithm auditing represents only a single facet within the broader algorithm accountability framework. To establish genuine algorithm accountability, it is imperative to consider that algorithm auditing operates with various other factors, forming an intricate ecosystem \cite{ada2020examining}. This is why we demand the AI audit ecosystem to be open and interfere with other accountability processes. An essential component for that is actual enforcement, that is, post-audit actions such as \citet{raji_actionable_2019}'s demands for harm reduction. This in turn requires European institutions that implement proper enforcement.

\section{Addressing the Regulatory Gap: An EU AI Audit Ecosystem}
\label{sec:gap}

Third-party audits by research and civil society are essential for an AI audit ecosystem. (1) Third parties can provide oversight as watchdogs against audit washing and false assurances. (2) Third-party audits can question standards, which could prevent the decoupling of audits, as in financial audits. (3) Third-party audits are adaptive to new technologies. In the context of (2) and (3), conducting audits through third-party entities gains further significance, as they can scrutinize evolving standards and raise critical inquiries \cite{laux_three_2023}. (4) Third parties promote sensitivity to the sociotechnical system and (5) affected communities. (6) Last but not least, third-party audits and data access challenge information and power asymmetries of the platform economy. At the same time, we have shown that the DSA and AIA refer differently to third-party audits and permit different data access for third parties. We will assess these approaches to third parties and propose policy recommendations. 

\subsection{Evaluation of the Audit Status Quo in the DSA and the AIA }
The DSA and the draft delegated act do not specify any additional requirements for the audit firm.\footnote{The wording of Art. 37 DSA (“the organisations carrying out the audits”, Art. 37 (2) DSA) does not provide any further details on this. Thus, Art. 4 (1) Delegated Regulation on Independent Audits refers to Art. 37 DSA, according to which the provider must check whether the organisation to be selected meets the requirements of paragraph 3 before selecting the audit organisation. Furthermore, the Commission defines the term audit organisation means an individual organisation, a consortium or other combination of organisations, including any sub-contractors, that the audited provider has contracted to perform an independent audit in accordance with Article 37 DSA.} This should be viewed critically for the 
two reasons: Firstly, it does not specify the requirements to be met by each criterion. For example, it does not explain in sufficient detail what level of “proven expertise” or supervisory steps are required to ensure compliance. In addition, the requirement of independence gives rise to discussion.\footnote{For example, \citet{Spindler+2022+689+704}, who raises concerns, for example, that this does not clarify constellations in which an affiliate of the auditing company advises the provider} Secondly, it is unclear whether audit firms need to be reviewed by bodies other than the VLOPs and VLOEs to ensure their proper functioning.  However, as a regulatory instrument can only be as effective as the institutions that enforce it, the final version of the delegated regulation needs to set out more specific requirements. On the one hand, it should give priority to independent checks on the competent selection of auditors by setting out basic independence requirements. On the other hand, the regulation should provide for additional (official) supervision.

As noted above, the EU AIA trilogue version relies primarily on internal conformity assessment as a means of assurance. The fact that the focus of the AIA is mainly on self-regulation is questionable because there is a lot at stake: high-risk AI systems covered by the AIA can violate numerous legal interests, e. g. people's health and lives at risk, but also damage property or the environment or violate legal interests such as the protection of privacy. Moreover, implementing the risk-based approach of the AIA requires taking preventive measures that should not solely rely on the assessment of the AI provider, especially for high-risk systems \cite{kop_eu_2021}. 

Given the risks, whether a first-party internal audit can provide the necessary oversight of such potentially harmful high-risk AI systems is questionable. Despite regulatory requirements, enforcing internal conformity assessments is likely to be less stringent than it appears at first glance. In addition, conformity assessments that the provider voluntarily has carried out by third parties are likely to be rare in practice. Indeed, providers of high-risk systems are usually in the best position to understand and assess the conformity of the AI systems they develop. However, it is doubtful that first parties will seek to find all potential harms and thus find out their system does not comply with the AIA's legal standards. We have laid out the issues of internal audits in the previous section.

For the above reasons, the requirements for audit firms in both the DSA and the AIA must be amended. The explicit inclusion of third parties would be beneficial under both acts. Under the current rules, there is a risk that providers will only engage second-party auditors to conduct mandatory annual assessments. This focus can not create an open, dynamic, and self-organizing ecosystem including various stakeholders. We have established the value of third-party audits for an AI audit ecosystem and shown that AI regulation in the EU must include third-party audits and their data access to account for appropriate oversight apart from compliance by first-party audits.  

\subsection{Advantages of Broader Data Access for (Vetted) Researchers}
Purely relying on process access for third-party audits, as exemplified by the AMS algorithm in section \ref{sec:access}, presents difficulties due to the inherent opacity of the system’s workings. This is why there is a growing call for a graduated system of access levels among various groups \cite{koshiyama_towards_2021}. Consequently, we advocate for a minimum requirement to enable 'black-box' access, preferably extended to non-vetted researchers without the need for formal accreditation. Reddit, Facebook, and Twitter opted to discontinue their APIs, which had been the authorized avenue for third parties to access and retrieve information regarding user activities on the platform. This situation exposes researchers to legal liabilities due to potential breaches of, for example, Facebook’s Terms of Service, as previously elucidated in the context of AlgorithmWatch’s data donation case \cite{mancosu}. 

The research efforts undertaken in this context should be safeguarded by robust safe-harbor provisions that grant legal immunity to these researchers thereby shielding them from potential legal charges linked to hacking or unauthorized access, as the AI Now Institute demands \cite{institute_algorithmic_2023}. This framework would offer protection to those conducting audits that potentially expose harm, thereby incentivizing such assessments. More widespread access to research data can, to some extent, reduce this power imbalance through ''audit capture'' \cite{laux_taming_2021}. Broader access to data should prevent such scenarios of audit capture, seen when Facebook halted AlgorithmWatch’s research on Instagram; this emphasizes the role of third-party audits and other measures to reduce market power. 

Consequently, one fundamental proposition is the establishment of a right to scrape, although an even more effective approach could involve mandatory APIs. However, implementing the latter may pose challenges for non-vetted researchers, as it might inadvertently allow them to utilize the product without incurring costs. For vetted researchers, broader access should be granted, enabling performance assessments akin to the approach taken in the Gender Shades case and possibly allowing the measurement of facial recognition performance. We acknowledge that the DSA includes such access for vetted researchers. 

We have seen that NGOs have can ensure social accountability and oversight trough domain knowledge and inclusion of affected people. At the same time, journalists have already shown in the past how important they are as an accountability component in financial audits and also in algorithmic audits. It is for this reason that we believe that the extension of vetted researchers to include NGOs and journalists makes sense. This extension could be done in the delegated acts by describing vetted researchers in a broad manner such that  NGOs and journalists are included.

As has been shown, the AIA does not contain any provisions comparable to the DSA that facilitate regulated access by researchers to data held by AI providers. This is surprising given that, according to the rationale for its adoption, the AIA should in principle facilitate this. We recommend that this position be reconsidered and added to the final document.

\subsection{Policy Recommendations}

\paragraph{Explicit Inclusion of Third Parties}
Future AI regulations should explicitly include provisions for third-party audits and third-party access by researchers, civil society, and regulators. This ensures a more comprehensive and independent assessment of AI products and platform and moves towards establishing an EU AI audit ecosystem.

\paragraph{Inclusion of NGOs and Investigative Journalists}
In the DSA delegated act on data access, we recommend including NGOs and investigative journalists as vetted researchers who can perform third-party audits. Their involvement can ensure social accountability through domain knowledge and special access to affected communities. Additionally, journalists have demonstrated their importance in financial and algorithmic audits as an accountability component in the past.

\paragraph{Third-Party Access in AIA}
Incorporate provisions for third-party access in the AIA, particularly for ``high-risk'' AI systems. These provisions ensure that critical systems undergo rigorous scrutiny and that hard standards can be questioned in the light of new technologies.

\paragraph{Incentivize Third-Party Audits}
Create incentives, such as funding or tax benefits, to encourage organizations to undergo third-party audits voluntarily. These incentives can be funded through companies managed by the government or government funding. This promotes an enabling environment for an AI audit ecosystem and thus the promotion of algorithmic conformity assessment procedures within the framework of all regulations. The government could choose the researchers eligible for funding in the case of the DSA (should be thought of in the delegated acts) and through a multistakeholder board in the AIA which includes state-run ADMs. 

\paragraph{Publish Audit Results}
Nondisclosures of audit results should be an exception if sensitive information or trade secrets are undeniably part of the audit results. 
Otherwise, mandate the publication of anonymized audit results, whether conducted by first, second, or third parties. This can inform other auditors and promote further knowledge sharing.

\paragraph{Harm Incident Reporting and Detection}
Implement measures to detect and report AI-related harms. Establish a framework for reporting incidents and ensure post-audit actions and enforcement are taken.

\paragraph{Whistleblower Protections in AI Companies}
Whistleblower protections for employees who expose violations of AI standards, legal and regulatory requirements, civil rights, and human rights should be strengthened. Whistleblowers should be shielded from retaliation. The EU Whistleblower Protection Directive \cite{eu-directive-2019-1937} is a step in the right direction and will be welcomed. It will be compelling to see how AIA and the EU Whistleblower Protection Directive interact.

\paragraph{Safe Harbor Provisions for Auditing Entities}
Implement safe harbor provisions specifically for researchers, NGOs, and investigative journalists who conduct AI audits. These entities should be granted legal protection from retaliatory actions when conducting audits in good faith. Safe harbor provisions should apply when audits are conducted according to established standards and guidelines and findings are reported accurately.

\paragraph{Legalization of Web Scraping for AI Auditing}
Recognize the importance of web scraping as a valuable tool for researchers, NGOs, and investigative journalists engaged in AI auditing activities. Establish clear legal frameworks that permit responsible web scraping when it is carried out for legitimate research and auditing purposes. These frameworks should include safeguards to protect against misuse and ensure the ethical and responsible use of web scraping techniques.

\section{Conclusion}
We defined audit types -- focusing on third-party audits -- and analyzed to which extent they are included in the EU regulatory framework, with a focus on the DSA and the AIA. 

According to our analysis, the DSA will establish a novel institutional system that involves third-party audits,   provides data access for vetted researchers as well as access to openly available data for non-vetted researchers. However, the requirements for independence of auditors in the DSA still need to be specified. The AIA, on its turn, relies primarily on internal conformity assessment with external assessments that can be mandated by the regulator. At present, access for researchers or independent auditors is not explicitly intended.

Based on this analysis, we define an AI audit ecosystem that accounts for compliance and oversight and demonstrate the existence of a regulatory gap. We emphasize that third-party audits by independent organisations, researchers and civil society, as well as through inspection by regulators, must be part of that ecosystem. The AIA, in this sense, should have included data and model access for vetted researchers and civil society organisations to assess ``high-risk'' AI products, as is partially the case of the DSA's provisions allowing for vetted researchers — although not civil society organisations — to have access to sensitive information on the AI systems.

We hope to contribute to the emerging field of audits as a means of accountability. In doing so, we aim to sharpen the focus on affected communities and incorporate past lessons from platform research and financial audits. We recommend establishing a diverse AI audit ecosystem to ensure compliance and oversight. We concur with \citet{raji_outsider_2022}'s demands for an ``ecosystem in which third parties can not only survive in their role, but thrive in directly confronting, verifying, and subjecting to scrutiny the performance claims made by corporations, while adequately addressing complaints of harm brought forth by the impacted population.'' Our policy proposals point to democratizing accountability, and we hope future AI regulations and the EU directives mentioned here will address them.


\section{Compliance with Ethical Standards}
\subsection{Conflict of Interest} On behalf of all authors, the corresponding author states that there is no conflict of interest.

%
%
%
 \bibliographystyle{splncs04nat}
  \bibliography{references}

\begin{thebibliography}{103}
\providecommand{\natexlab}[1]{#1}
\providecommand{\url}[1]{\texttt{#1}}
\providecommand{\urlprefix}{URL }
\expandafter\ifx\csname urlstyle\endcsname\relax
  \providecommand{\doi}[1]{doi:\discretionary{}{}{}#1}\else
  \providecommand{\doi}{doi:\discretionary{}{}{}\begingroup \urlstyle{rm}\Url}\fi

\bibitem[{Ackerman(2005)}]{ackerman2005social}
Ackerman, J.M.: Social accountability in the public sector. Social Development Papers  (2005)

\bibitem[{{Ada Lovelace Institute}(2020)}]{ada2020examining}
{Ada Lovelace Institute}: Examining the black box: Tools for assessing algorithmic systems. the Black Box  (2020), \urlprefix\url{https://www.adalovelaceinstitute.org/report/examining-the-black-box-tools-for-assessing-algorithmic-systems/}

\bibitem[{Albert(2023)}]{albert_john_diverse_2023}
Albert, J.: A diverse auditing ecosystem is needed to uncover algorithmic risks (2023), \urlprefix\url{https://algorithmwatch.org/en/diverse-auditing-ecosystem-for-algorithmic-risks/}

\bibitem[{Algorithmwatch(2021)}]{Algorithmwatch2021}
Algorithmwatch: Draft ai act: {EU} needs to live up to its own ambitions in terms of governance and enforcement (2021), \urlprefix\url{https://algorithmwatch.org/en/wp-content/uploads/2021/08/EU-AI-Act-Consultation-Submission-by-AlgorithmWatch-August-2021.pdf}, accessed on August 3, 2023

\bibitem[{Allhutter et~al.(2020)Allhutter, Mager, Cech, Fischer, and Grill}]{allhutter2020ams}
Allhutter, D., Mager, A., Cech, F., Fischer, F., Grill, G.: Der ams algorithmus - eine soziotechnische analyse des arbeitsmarktchancen-assistenz-systems (amas). Tech. Rep. ITA 2020-02, Institut für Technikfolgen-Abschätzung der Österreichischen Akademie der Wissenschaften, Wien (2020)

\bibitem[{Asghari et~al.(2022)Asghari, Birner, Burchardt, Dicks, Faßbender, Feldhus, Hewett, Hofmann, Kettemann, Schulz, Simon, Stolberg-Larsen, and Züger}]{asghari_what_2022}
Asghari, H., Birner, N., Burchardt, A., Dicks, D., Faßbender, J., Feldhus, N., Hewett, F., Hofmann, V., Kettemann, M.C., Schulz, W., Simon, J., Stolberg-Larsen, J., Züger, T.: What to explain when explaining is difficult. {An} interdisciplinary primer on {XAI} and meaningful information in automated decision-making. Tech. rep., Zenodo (Mar 2022), \urlprefix\url{https://zenodo.org/record/6375784}

\bibitem[{Bail et~al.(2018)Bail, Argyle, Brown, Bumpus, Chen, Hunzaker, Lee, Mann, Merhout, and Volfovsky}]{Bail.2018}
Bail, C.A., Argyle, L.P., Brown, T.W., Bumpus, J.P., Chen, H., Hunzaker, M.B.F., Lee, J., Mann, M., Merhout, F., Volfovsky, A.: Exposure to opposing views on social media can increase political polarization. Proceedings of the National Academy of Sciences of the United States of America \textbf{115}(37), 9216--9221 (2018), ISSN 0027-8424

\bibitem[{Bandy(2021)}]{bandy_problematic_2021}
Bandy, J.: Problematic {Machine} {Behavior}: {A} {Systematic} {Literature} {Review} of {Algorithm} {Audits}. Proceedings of the ACM on Human-Computer Interaction \textbf{5}(CSCW1), 74:1--74:34 (Apr 2021), \urlprefix\url{https://doi.org/10.1145/3449148}

\bibitem[{Bazerman et~al.(1997)Bazerman, Morgan, Loewenstein et~al.}]{bazerman1997impossibility}
Bazerman, M.H., Morgan, K.P., Loewenstein, G.F., et~al.: The impossibility of auditor independence. Sloan management review \textbf{38}, 89--94 (1997)

\bibitem[{Berthin(2011)}]{berthin}
Berthin, G.: A practical guide to social audit as a participatory tool to strengthen democratic governance, transparency, and accountability (09 2011), \urlprefix\url{https://www.undp.org/sites/g/files/zskgke326/files/migration/latinamerica/Practical-Guide-to-Social-Audit.pdf}

\bibitem[{Bieker et~al.(2016)Bieker, Friedewald, Hansen, Obersteller, and Rost}]{bieker_process_2016}
Bieker, F., Friedewald, M., Hansen, M., Obersteller, H., Rost, M.: A {Process} for {Data} {Protection} {Impact} {Assessment} {Under} the {European} {General} {Data} {Protection} {Regulation}. In: Schiffner, S., Serna, J., Ikonomou, D., Rannenberg, K. (eds.) Privacy {Technologies} and {Policy}, pp. 21--37, Lecture {Notes} in {Computer} {Science}, Springer International Publishing, Cham (2016), ISBN 978-3-319-44760-5

\bibitem[{Birhane et~al.(2024)Birhane, Steed, Ojewale, Vecchione, and Raji}]{birhane_ai_2024}
Birhane, A., Steed, R., Ojewale, V., Vecchione, B., Raji, I.D.: {AI} auditing: {The} {Broken} {Bus} on the {Road} to {AI} {Accountability} (Jan 2024), \urlprefix\url{http://arxiv.org/abs/2401.14462}, arXiv:2401.14462 [cs]

\bibitem[{Brown et~al.(2021)Brown, Davidovic, and Hasan}]{brown}
Brown, S., Davidovic, J., Hasan, A.: The algorithm audit: Scoring the algorithms that score us. Big Data \& Society \textbf{8}(1), 2053951720983865 (2021), \urlprefix\url{https://doi.org/10.1177/2053951720983865}

\bibitem[{Buolamwini and Gebru(2018)}]{buolamwini_gender_nodate}
Buolamwini, J., Gebru, T.: Gender shades: Intersectional accuracy disparities in commercial gender classification. In: Conference on fairness, accountability and transparency, pp. 77--91, PMLR (2018)

\bibitem[{Buri and van Hoboken(2021)}]{buri-van-hoboken-dsa-observatory}
Buri, J., van Hoboken, J.: The digital services act (dsa) proposal: a critical overview – discussion paper. Discussion Paper 28 October 2021, Digital Services Act (DSA) Observatory (2021)

\bibitem[{Casper et~al.(2024)Casper, Ezell, Siegmann, Kolt, Curtis, Bucknall, Haupt, Wei, Scheurer, Hobbhahn, Sharkey, Krishna, Von~Hagen, Alberti, Chan, Sun, Gerovitch, Bau, Tegmark, Krueger, and Hadfield-Menell}]{casper_black-box_2024}
Casper, S., Ezell, C., Siegmann, C., Kolt, N., Curtis, T.L., Bucknall, B., Haupt, A., Wei, K., Scheurer, J., Hobbhahn, M., Sharkey, L., Krishna, S., Von~Hagen, M., Alberti, S., Chan, A., Sun, Q., Gerovitch, M., Bau, D., Tegmark, M., Krueger, D., Hadfield-Menell, D.: Black-{Box} {Access} is {Insufficient} for {Rigorous} {AI} {Audits} (Jan 2024), \urlprefix\url{http://arxiv.org/abs/2401.14446}, arXiv:2401.14446 [cs]

\bibitem[{Castets-Renard and Besse(ming)}]{castets-renard-ai-act}
Castets-Renard, C., Besse, P.: Ex ante accountability of the ai act: Between certification and standardization. Artificial Intelligence Law: Between Sectoral Rules and Comprehensive Regime. Comparative Law Perspectives  (Forthcoming), \urlprefix\url{https://ssrn.com/abstract=4203925}

\bibitem[{Chen et~al.(2018)Chen, Chiou, Huang, Tu, Lee, and Chien}]{doi:10.1177/2168479017716712}
Chen, Y.J., Chiou, C.M., Huang, Y.W., Tu, P.W., Lee, Y.C., Chien, C.H.: A comparative study of medical device regulations:: Us, europe, canada, and taiwan. Therapeutic Innovation \& Regulatory Science \textbf{52}(1), 62--69 (2018), \urlprefix\url{https://doi.org/10.1177/2168479017716712}, pMID: 29714608

\bibitem[{Commission(2023)}]{dsa-audit}
Commission, E.: Draft for commission delegated regulation (eu) by laying down rules on the performance of audits for very large online platforms and very large online search engines (2023), \urlprefix\url{missing}, 5 May 2023

\bibitem[{Constantaras et~al.(2023{\natexlab{a}})Constantaras, Geiger, Braun, Mehrotra, and Aung}]{wire}
Constantaras, E., Geiger, G., Braun, J.C., Mehrotra, D., Aung, H.: Inside the {Suspicion} {Machine}. Wired  (2023{\natexlab{a}}), ISSN 1059-1028, \urlprefix\url{https://www.wired.com/story/welfare-state-algorithms/}, section: tags

\bibitem[{Constantaras et~al.(2023{\natexlab{b}})Constantaras, Geiger, Braun, Mehrotra, and Aung}]{noauthor_suspicion_nodate}
Constantaras, E., Geiger, G., Braun, J.C., Mehrotra, D., Aung, H.: Suspicion {Machines} {Methodology} - {Lighthouse} {Reports} (2023{\natexlab{b}}), \urlprefix\url{https://www.lighthousereports.com/suspicion-machines-methodology/}

\bibitem[{Costanza-Chock et~al.(2022)Costanza-Chock, Raji, and Buolamwini}]{costanza-chock_who_2022}
Costanza-Chock, S., Raji, I.D., Buolamwini, J.: Who {Audits} the {Auditors}? {Recommendations} from a field scan of the algorithmic auditing ecosystem. In: 2022 {ACM} {Conference} on {Fairness}, {Accountability}, and {Transparency}, pp. 1571--1583, ACM, Seoul Republic of Korea (Jun 2022), ISBN 978-1-4503-9352-2, \urlprefix\url{https://dl.acm.org/doi/10.1145/3531146.3533213}

\bibitem[{Ebers et~al.(2021)Ebers, Hoch, Rosenkranz, Ruschemeier, and Steinrötter}]{j4040043}
Ebers, M., Hoch, V.R.S., Rosenkranz, F., Ruschemeier, H., Steinrötter, B.: The european commission’s proposal for an artificial intelligence act—a critical assessment by members of the robotics and ai law society (rails). J \textbf{4}(4), 589--603 (2021), ISSN 2571-8800, \urlprefix\url{https://www.mdpi.com/2571-8800/4/4/43}

\bibitem[{Edwards(2022)}]{edwards2022}
Edwards, L.: Regulating ai in europe: four problems and four solutions (2022), \urlprefix\url{https://www.adalovelaceinstitute.org/report/regulating-ai-in-europe/}

\bibitem[{{European Commision}(2021)}]{AIA}
{European Commision}: Proposal for a regulation of the european parliament and of the council laying down harmonised rules on artificial intelligence (artificial intelligence act) and amending certain union legislative acts (2021), \urlprefix\url{https://eur-lex.europa.eu/legal-content/EN/TXT/?uri=celex%3A52021PC0206}, 18.08.23

\bibitem[{{European Commission}(2022)}]{european_commission_dsa_nodate}
{European Commission}: {DSA}: {Very} {Large} {Online} {Platforms} and {Search} {Engines} (2022), \urlprefix\url{https://ec.europa.eu/commission/presscorner/detail/en/IP_23_2413}, press release No. IP/23/2413 from 25.04.23

\bibitem[{{European Commission}(2023)}]{euregulation}
{European Commission}: Commission delegated regulation (eu) …/… of 20.10.2023 supplementing regulation (eu) 2022/2065 of the european parliament and of the council, by laying down rules on the performance of audits for very large online platforms and very large online search engines (October 20 2023), \urlprefix\url{https://eur-lex.europa.eu/legal-content/EN/TXT/?uri=CELEX%3A32023R6807}

\bibitem[{{European Parliament}(2019)}]{eu-directive-2019-1937}
{European Parliament}: Directive (eu) 2019/1937 of the european parliament and of the council. Directive EU 2019/1937, European Parliament (October 2019)

\bibitem[{{European Parliament}(2022)}]{EU-AI-Report}
{European Parliament}: Draft report on the proposal for a regulation of the european parliament and of the council on harmonised rules on artificial intelligence (artificial intelligence act) and amending certain union legislative acts. Draft Report COM2021/0206 – C9-0146/2021 – 2021/0106(COD), Committee on the Internal Market and Consumer Protection, Committee on Civil Liberties, Justice and Home Affairs (April 20 2022)

\bibitem[{{European Parliament}(2024)}]{AIAct2024}
{European Parliament}: Corrigendum to the position of the european parliament adopted at first reading on 13 march 2024 with a view to the adoption of regulation (eu) 2024/ ...... of the european parliament and of the council laying down harmonised rules on artificial intelligence and amending regulations (ec) no 300/2008, (eu) no 167/2013, (eu) no 168/2013, (eu) 2018/858, (eu) 2018/1139 and (eu) 2019/2144 and directives 2014/90/eu, (eu) 2016/797 and (eu) 2020/1828 (artificial intelligence act). Available online: \url{https://www.europarl.europa.eu/doceo/document/TA-9-2024-0138-FNL-COR01_EN.pdf} (2024), position of the European Parliament adopted at first reading on 13 March 2024. P9\_TA(2024)0138 (COM(2021)0206 - C9-0146/2021 - 2021/0106(COD))

\bibitem[{{European Parliament} and {Council of the European Union}(2022)}]{dsa}
{European Parliament}, {Council of the European Union}: Regulation ({EU}) 2022/2065 of the {European} {Parliament} and of the {Council} of 19 {October} 2022 on a {Single} {Market} {For} {Digital} {Services} and amending {Directive} 2000/31/{EC} ({Digital} {Services} {Act}) (2022), \urlprefix\url{https://eur-lex.europa.eu/eli/reg/2022/2065/oj}

\bibitem[{Floridi et~al.(2022)Floridi, Holweg, Taddeo, Amaya~Silva, Mökander, and Wen}]{floridi_capai_2022}
Floridi, L., Holweg, M., Taddeo, M., Amaya~Silva, J., Mökander, J., Wen, Y.: {capAI} - {A} {Procedure} for {Conducting} {Conformity} {Assessment} of {AI} {Systems} in {Line} with the {EU} {Artificial} {Intelligence} {Act} (Mar 2022), \urlprefix\url{https://papers.ssrn.com/abstract=4064091}

\bibitem[{Gaddis(2017)}]{gaddis_introduction_2017}
Gaddis, S.M.: An {Introduction} to {Audit} {Studies} in the {Social} {Sciences} (Aug 2017), \urlprefix\url{https://papers.ssrn.com/abstract=3024262}

\bibitem[{Globus-Harris et~al.(2022)Globus-Harris, Kearns, and Roth}]{globus-harris_algorithmic_2022}
Globus-Harris, I., Kearns, M., Roth, A.: An {Algorithmic} {Framework} for {Bias} {Bounties} (May 2022), \urlprefix\url{http://arxiv.org/abs/2201.10408}, arXiv:2201.10408 [cs]

\bibitem[{Goldhaber-Fiebert and Prince(2019)}]{goldhaber2019impact}
Goldhaber-Fiebert, J.D., Prince, L.: Impact evaluation of a predictive risk modeling tool for allegheny county’s child welfare office (March 20 2019), \urlprefix\url{https://www.alleghenycountyanalytics.us/wp-content/uploads/2019/05/Impact-Evaluation-from-16-ACDHS-26_PredictiveRisk_Package_050119_FINAL-6.pdf}

\bibitem[{Goodman and Trehu(2022)}]{goodman_ai_2022}
Goodman, E.P., Trehu, J.: {AI} {Audit} {Washing} and {Accountability} (Sep 2022), \urlprefix\url{https://papers.ssrn.com/abstract=4227350}

\bibitem[{Groves(2022)}]{groves_algorithmic_nodate}
Groves, L.: Algorithmic impact assessment: a case study in healthcare (2022), \urlprefix\url{https://www.adalovelaceinstitute.org/report/algorithmic-impact-assessment-case-study-healthcare/}

\bibitem[{Halloran(2021)}]{halloran2021accountability}
Halloran, B.: Accountability ecosystems: The evolution of a keyword. Accountability Research Center  (2021)

\bibitem[{Hansen and Loftus(2023)}]{hansen_model-agnostic_nodate}
Hansen, S., Loftus, J.: Model-{Agnostic} {Auditing}: {A} {Lost} {Cause}? Proceedings of the EWAF’23: European Workshop on Algorithmic Fairness  (2023)

\bibitem[{Heuer(2021)}]{heuer_audit_2021}
Heuer, H.: Audit, don’t explain – recommendations based on a socio-technical understanding of ml-based systems. Mensch und Computer 2021 - Workshopband  (2021)

\bibitem[{Hickey et~al.(2023)Hickey, Schmitz, Fessler, Smaldino, Muric, and Burghardt}]{hickey2023auditing}
Hickey, D., Schmitz, M., Fessler, D., Smaldino, P.E., Muric, G., Burghardt, K.: Auditing elon musk’s impact on hate speech and bots. In: Proceedings of the International AAAI Conference on Web and Social Media, vol.~17, pp. 1133--1137 (2023)

\bibitem[{Kaesling(2023)}]{kaesling-dsa}
Kaesling: Art. 40. In: Hofmann, Raue (eds.) Digital Service Act, p.~33, Nomos (2023)

\bibitem[{Kak and West(2023)}]{institute_algorithmic_2023}
Kak, A., West, S.M.: Algorithmic {Accountability}: {Moving} {Beyond} {Audits}. AI Now Institute  (Apr 2023), \urlprefix\url{https://ainowinstitute.org/publication/algorithmic-accountability}

\bibitem[{Kayser-Bril(2021)}]{instagramalgorithmwatch}
Kayser-Bril, N.: {AlgorithmWatch} forced to shut down {Instagram} monitoring project after threats from {Facebook} (2021), \urlprefix\url{https://algorithmwatch.org/en/instagram-research-shut-down-by-facebook/}

\bibitem[{Kazim et~al.(2021)Kazim, Denny, and Koshiyama}]{kazim2021ai}
Kazim, E., Denny, D.M., Koshiyama, A.: Ai auditing and impact assessment: According to the uk information commissioner's office. AI Ethics \textbf{1}, 301--310 (2021), \urlprefix\url{https://doi.org/10.1007/s43681-021-00039-2}

\bibitem[{Kop(2021)}]{kop_eu_2021}
Kop, M.: {EU} {Artificial} {Intelligence} {Act}: {The} {European} {Approach} to {AI} (Sep 2021), \urlprefix\url{https://papers.ssrn.com/abstract=3930959}

\bibitem[{Koshiyama et~al.(2022)Koshiyama, Kazim, and Treleaven}]{9755237}
Koshiyama, A., Kazim, E., Treleaven, P.: Algorithm auditing: Managing the legal, ethical, and technological risks of artificial intelligence, machine learning, and associated algorithms. Computer \textbf{55}(4), 40--50 (2022), \doi{10.1109/MC.2021.3067225}

\bibitem[{Koshiyama et~al.(2021)Koshiyama, Kazim, Treleaven, Rai, Szpruch, Pavey, Ahamat, Leutner, Goebel, Knight, Adams, Hitrova, Barnett, Nachev, Barber, Chamorro-Premuzic, Klemmer, Gregorovic, Khan, and Lomas}]{koshiyama_towards_2021}
Koshiyama, A., Kazim, E., Treleaven, P., Rai, P., Szpruch, L., Pavey, G., Ahamat, G., Leutner, F., Goebel, R., Knight, A., Adams, J., Hitrova, C., Barnett, J., Nachev, P., Barber, D., Chamorro-Premuzic, T., Klemmer, K., Gregorovic, M., Khan, S., Lomas, E.: Towards {Algorithm} {Auditing}: {A} {Survey} on {Managing} {Legal}, {Ethical} and {Technological} {Risks} of {AI}, {ML} and {Associated} {Algorithms}. SSRN  (Jan 2021), \urlprefix\url{https://papers.ssrn.com/abstract=3778998}

\bibitem[{Larson et~al.(2016)Larson, Kirchner, Mattu, and Angwin}]{mattu_how_nodate}
Larson, J., Kirchner, L., Mattu, S., Angwin, J.: How {We} {Analyzed} the {COMPAS} {Recidivism} {Algorithm} (2016), \urlprefix\url{https://www.propublica.org/article/how-we-analyzed-the-compas-recidivism-algorithm}

\bibitem[{Laux et~al.(2021)Laux, Wachter, and Mittelstadt}]{laux_taming_2021}
Laux, J., Wachter, S., Mittelstadt, B.: Taming the {Few}: {Platform} {Regulation}, {Independent} {Audits}, and the {Risks} of {Capture} {Created} by the {DMA} and {DSA} (Sep 2021), \urlprefix\url{https://papers.ssrn.com/abstract=4096655}

\bibitem[{Laux et~al.(2023)Laux, Wachter, and Mittelstadt}]{laux_three_2023}
Laux, J., Wachter, S., Mittelstadt, B.: Three {Pathways} for {Standardisation} and {Ethical} {Disclosure} by {Default} under the {European} {Union} {Artificial} {Intelligence} {Act}. SSRN Electronic Journal  (2023), ISSN 1556-5068, \urlprefix\url{https://www.ssrn.com/abstract=4365079}

\bibitem[{Laux et~al.(2024)Laux, Wachter, and Mittelstadt}]{laux_three_2024}
Laux, J., Wachter, S., Mittelstadt, B.: Three pathways for standardisation and ethical disclosure by default under the {European} {Union} {Artificial} {Intelligence} {Act}. Computer Law \& Security Review \textbf{53}, 105957 (2024), ISSN 0267-3649, \urlprefix\url{https://www.sciencedirect.com/science/article/pii/S0267364924000244}

\bibitem[{Li et~al.(2023)Li, Kingsley, Fan, Sinha, Wai, Lee, Shen, Eslami, and Hong}]{li}
Li, R., Kingsley, S., Fan, C., Sinha, P., Wai, N., Lee, J., Shen, H., Eslami, M., Hong, J.: Participation and division of labor in user-driven algorithm audits: How do everyday users work together to surface algorithmic harms? CHI '23: Proceedings of the 2023 CHI Conference on Human Factors in Computing Systems  (2023)

\bibitem[{Löhlein and Huber(2022)}]{lohlein-huber-2022}
Löhlein, L., Huber, C.: The end of audit: Spectacle and love in the audit society. Qualitative Research in Accounting \& Management  (2022)

\bibitem[{Mancosu and Vegetti(2020)}]{mancosu}
Mancosu, M., Vegetti, F.: What you can scrape and what is right to scrape: A proposal for a tool to collect public facebook data. Social Media + Society \textbf{6}(3), 2056305120940703 (2020), \urlprefix\url{https://doi.org/10.1177/2056305120940703}

\bibitem[{McCrum(2015)}]{ft-wirecard}
McCrum, D.: The house of wirecard (2015), \urlprefix\url{https://www.ft.com/content/534e7c4d-3101-3f6a-abc8-dc70beab35b7}, financial Times, April 27, 2015

\bibitem[{McFadden et~al.(2021)McFadden, Jones, Taylor, and Osborn}]{mcfadden_harmonising_nodate}
McFadden, M., Jones, K., Taylor, E., Osborn, G.: Harmonising {Artificial} {Intelligence:} the role of standards in the eu ai regulation. Oxford Internet Institute  (2021)

\bibitem[{McNamara et~al.(2018)McNamara, Smith, and Murphy-Hill}]{mcnamara_does_2018}
McNamara, A., Smith, J., Murphy-Hill, E.: Does {ACM}’s code of ethics change ethical decision making in software development? In: Proceedings of the 2018 26th {ACM} {Joint} {Meeting} on {European} {Software} {Engineering} {Conference} and {Symposium} on the {Foundations} of {Software} {Engineering}, pp. 729--733, ACM, Lake Buena Vista FL USA (Oct 2018), ISBN 978-1-4503-5573-5, \urlprefix\url{https://dl.acm.org/doi/10.1145/3236024.3264833}

\bibitem[{Metaxa et~al.(2021)Metaxa, Park, Robertson, Karahalios, Wilson, Hancock, and Sandvig}]{HCI-083}
Metaxa, D., Park, J.S., Robertson, R.E., Karahalios, K., Wilson, C., Hancock, J., Sandvig, C.: Auditing algorithms: Understanding algorithmic systems from the outside in. Foundations and Trends® in Human–Computer Interaction \textbf{14}(4), 272--344 (2021), ISSN 1551-3955, \urlprefix\url{http://dx.doi.org/10.1561/1100000083}

\bibitem[{Meßmer and Degeling(2023)}]{mesmer_auditing_2023}
Meßmer, A.K., Degeling, M.: Auditing {Recommender} {Systems} (Jan 2023), \urlprefix\url{https://www.stiftung-nv.de/de/publication/auditing-recommender-systems}

\bibitem[{Munn(2023)}]{munn_uselessness_2023}
Munn, L.: The uselessness of {AI} ethics. AI and Ethics \textbf{3}(3), 869--877 (Aug 2023), ISSN 2730-5953, 2730-5961, \urlprefix\url{https://link.springer.com/10.1007/s43681-022-00209-w}

\bibitem[{Mökander et~al.(2022)Mökander, Axente, Casolari, and Floridi}]{mokander_conformity_2022}
Mökander, J., Axente, M., Casolari, F., Floridi, L.: Conformity {Assessments} and {Post}-market {Monitoring}: {A} {Guide} to the {Role} of {Auditing} in the {Proposed} {European} {AI} {Regulation}. Minds and Machines \textbf{32}(2), 241--268 (Jun 2022), ISSN 1572-8641, \urlprefix\url{https://doi.org/10.1007/s11023-021-09577-4}

\bibitem[{Mökander and Floridi(2021)}]{mokander_ethics-based_2021}
Mökander, J., Floridi, L.: Ethics-{Based} {Auditing} to {Develop} {Trustworthy} {AI}. Minds and Machines \textbf{31}(2), 323--327 (Jun 2021), ISSN 1572-8641, \urlprefix\url{https://doi.org/10.1007/s11023-021-09557-8}

\bibitem[{Mökander et~al.(2023)Mökander, Schuett, Kirk, and Floridi}]{mokander_auditing_2023}
Mökander, J., Schuett, J., Kirk, H.R., Floridi, L.: Auditing large language models: a three-layered approach (Feb 2023), \urlprefix\url{http://arxiv.org/abs/2302.08500}, arXiv:2302.08500 [cs]

\bibitem[{Nicoletti and Bass(2023)}]{bloomberg}
Nicoletti, L., Bass, D.: Generative {AI} {Takes} {Stereotypes} and {Bias} {From} {Bad} to {Worse} (2023), \urlprefix\url{https://www.bloomberg.com/graphics/2023-generative-ai-bias/}

\bibitem[{Novelli et~al.(ming)Novelli, Taddeo, and Floridi}]{NovelliForthcoming-NOVAIA}
Novelli, C., Taddeo, M., Floridi, L.: Accountability in artificial intelligence: What it is and how it works. Ai and Society: Knowledge, Culture and Communication pp. 1--12 (forthcoming)

\bibitem[{Ohme et~al.(2023)Ohme, Araujo, Boeschoten, Freelon, Ram, Reeves, and Robinson}]{ohme_digital_2023}
Ohme, J., Araujo, T., Boeschoten, L., Freelon, D., Ram, N., Reeves, B.B., Robinson, T.N.: Digital {Trace} {Data} {Collection} for {Social} {Media} {Effects} {Research}: {APIs}, {Data} {Donation}, and ({Screen}) {Tracking}. Communication Methods and Measures pp. 1--18 (Feb 2023), ISSN 1931-2458, 1931-2466, \urlprefix\url{https://www.tandfonline.com/doi/full/10.1080/19312458.2023.2181319}

\bibitem[{Olteanu et~al.(2019)Olteanu, Castillo, Diaz, and K{\i}c{\i}man}]{olteanu2019social}
Olteanu, A., Castillo, C., Diaz, F., K{\i}c{\i}man, E.: Social data: Biases, methodological pitfalls, and ethical boundaries. Frontiers in big data \textbf{2}, 13 (2019)

\bibitem[{Platform(2022)}]{twitterapi}
Platform, X.D.: Changelog {\textbar} {Twitter} {Developer} {Platform} (2022), \urlprefix\url{https://developer.twitter.com/en/updates/changelog}

\bibitem[{Power(1999)}]{Power1999-POWTAS-3}
Power, M.: The audit society: Rituals of verification. British Journal of Educational Studies \textbf{47}(1), 92--94 (1999)

\bibitem[{Quattrociocchi(2017)}]{quattrociocchi_social_2017}
Quattrociocchi, W.: Social and political challenges: Western democracy in crisis? In: Global Risks Report 2017 (2017)

\bibitem[{Radiya-Dixit and Neff(2023)}]{sociotechnical}
Radiya-Dixit, E., Neff, G.: A sociotechnical audit: Assessing police use of facial recognition. In: Proceedings of the 2023 ACM Conference on Fairness, Accountability, and Transparency, p. 1334–1346, FAccT '23, Association for Computing Machinery, New York, NY, USA (2023), ISBN 9798400701924, \urlprefix\url{https://doi.org/10.1145/3593013.3594084}

\bibitem[{Raji(2022)}]{noauthor_mozilla_nodate}
Raji, D.: Mozilla {Open} {Source} {Audit} {Tooling} ({OAT}) {Project} (2022), \urlprefix\url{https://foundation.mozilla.org/en/what-we-fund/fellowships/oat/}

\bibitem[{Raji and Buolamwini(2019)}]{raji_actionable_2019}
Raji, I.D., Buolamwini, J.: Actionable {Auditing}: {Investigating} the {Impact} of {Publicly} {Naming} {Biased} {Performance} {Results} of {Commercial} {AI} {Products}. In: Proceedings of the 2019 {AAAI}/{ACM} {Conference} on {AI}, {Ethics}, and {Society}, pp. 429--435, ACM, Honolulu HI USA (Jan 2019), ISBN 978-1-4503-6324-2, \urlprefix\url{https://dl.acm.org/doi/10.1145/3306618.3314244}

\bibitem[{Raji et~al.(2020)Raji, Smart, White, Mitchell, Gebru, Hutchinson, Smith-Loud, Theron, and Barnes}]{raji2020closing}
Raji, I.D., Smart, A., White, R.N., Mitchell, M., Gebru, T., Hutchinson, B., Smith-Loud, J., Theron, D., Barnes, P.: Closing the ai accountability gap: Defining an end-to-end framework for internal algorithmic auditing. In: Proceedings of the 2020 conference on fairness, accountability, and transparency, pp. 33--44 (2020)

\bibitem[{Raji et~al.(2022)Raji, Xu, Honigsberg, and Ho}]{raji_outsider_2022}
Raji, I.D., Xu, P., Honigsberg, C., Ho, D.E.: Outsider {Oversight}: {Designing} a {Third} {Party} {Audit} {Ecosystem} for {AI} {Governance} (Jun 2022), \urlprefix\url{http://arxiv.org/abs/2206.04737}, arXiv:2206.04737 [cs]

\bibitem[{Rakova and Dobbe(2023)}]{rakova_algorithms_2023}
Rakova, B., Dobbe, R.: Algorithms as {Social}-{Ecological}-{Technological} {Systems}: an {Environmental} {Justice} {Lens} on {Algorithmic} {Audits}. In: 2023 {ACM} {Conference} on {Fairness}, {Accountability}, and {Transparency}, pp. 491--491 (Jun 2023), \urlprefix\url{http://arxiv.org/abs/2305.05733}, arXiv:2305.05733 [cs]

\bibitem[{Ray(2023)}]{ray_chatgpt_2023}
Ray, P.P.: {ChatGPT}: {A} comprehensive review on background, applications, key challenges, bias, ethics, limitations and future scope. Internet of Things and Cyber-Physical Systems \textbf{3}, 121--154 (2023), ISSN 26673452, \urlprefix\url{https://linkinghub.elsevier.com/retrieve/pii/S266734522300024X}

\bibitem[{Reddit(2023)}]{redditapi}
Reddit: Addressing the community about changes to our {API} : r/reddit (2023), \urlprefix\url{https://www.reddit.com/r/reddit/comments/145bram/addressing_the_community_about_changes_to_our_api/}

\bibitem[{Ribeiro et~al.(2020)Ribeiro, Ottoni, West, Almeida, and Meira}]{ribeiro20202}
Ribeiro, M.H., Ottoni, R., West, R., Almeida, V.A.F., Meira, W.: Auditing radicalization pathways on youtube. In: Proceedings of the 2020 Conference on Fairness, Accountability, and Transparency, p. 131–141, FAT* '20, Association for Computing Machinery, New York, NY, USA (2020), ISBN 9781450369367, \urlprefix\url{https://doi.org/10.1145/3351095.3372879}

\bibitem[{Sambasivan et~al.(2020)Sambasivan, Arnesen, Hutchinson, and Prabhakaran}]{sambasivan_non-portability_2020}
Sambasivan, N., Arnesen, E., Hutchinson, B., Prabhakaran, V.: Non-portability of {Algorithmic} {Fairness} in {India} (Dec 2020), \urlprefix\url{http://arxiv.org/abs/2012.03659}, arXiv:2012.03659 [cs]

\bibitem[{Sandvig et~al.(2014)Sandvig, Hamilton, Karahalios, and Langbort}]{sandvig_auditing_2014}
Sandvig, C., Hamilton, K., Karahalios, K., Langbort, C.: Auditing {Algorithms}: {Research} {Methods} for {Detecting} {Discrimination} on {Internet} {Platforms}. Data and Discrimination: Converting Critical Concerns into Productive Inquiry  (2014)

\bibitem[{Sartori and Theodorou(2022)}]{sartori2022sociotechnical}
Sartori, L., Theodorou, A.: A sociotechnical perspective for the future of ai: narratives, inequalities, and human control. Ethics and Information Technology \textbf{24}(4), 315--328 (2022), \doi{10.1007/s10676-022-09624-3}, \urlprefix\url{https://doi.org/10.1007/s10676-022-09624-3}

\bibitem[{Selinger and Hartzog(2016)}]{selinger_facebooks_2016}
Selinger, E., Hartzog, W.: Facebook’s emotional contagion study and the ethical problem of co-opted identity in mediated environments where users lack control. Research Ethics \textbf{12}(1), 35--43 (Jan 2016), ISSN 1747-0161, \urlprefix\url{https://journals.sagepub.com/doi/full/10.1177/1747016115579531}, publisher: SAGE Publications Ltd

\bibitem[{Shelby et~al.(2023)Shelby, Rismani, Henne, Moon, Rostamzadeh, Nicholas, Yilla-Akbari, Gallegos, Smart, Garcia, and Virk}]{shelby_sociotechnical_2022}
Shelby, R., Rismani, S., Henne, K., Moon, A., Rostamzadeh, N., Nicholas, P., Yilla-Akbari, N., Gallegos, J., Smart, A., Garcia, E., Virk, G.: Sociotechnical harms of algorithmic systems: Scoping a taxonomy for harm reduction. In: Proceedings of the 2023 AAAI/ACM Conference on AI, Ethics, and Society, p. 723–741, AIES '23, Association for Computing Machinery, New York, NY, USA (2023), ISBN 9798400702310, \urlprefix\url{https://doi.org/10.1145/3600211.3604673}

\bibitem[{Sikka(2009)}]{Sikka2009financial}
Sikka, P.: Financial crisis and the silence of the auditors. Accounting, Organizations and Society  (2009), \doi{10.1016/j.aos.2009.01.004}

\bibitem[{Spindler(2022)}]{Spindler+2022+689+704}
Spindler, G.: Die vorschläge der eu-kommission zu einer neuen produkthaftung und zur haftung von herstellern und betreibern künstlicher intelligenz. Computer und Recht \textbf{38}(11), 689--704 (2022), \urlprefix\url{https://doi.org/10.9785/cr-2022-381106}

\bibitem[{Stahl et~al.(2023)Stahl, Antoniou, Bhalla et~al.}]{stahl2023systematic}
Stahl, B.C., Antoniou, J., Bhalla, N., et~al.: A systematic review of artificial intelligence impact assessments. Artificial Intelligence Review  (2023)

\bibitem[{Sweeney(2013)}]{sweeney_discrimination_2013}
Sweeney, L.: Discrimination in {Online} {Ad} {Delivery}. SSRN Electronic Journal  (2013), ISSN 1556-5068, \urlprefix\url{http://www.ssrn.com/abstract=2208240}

\bibitem[{{The Ada Lovelace Institute}(2021)}]{the_ada_lovelace_institute_technical_2021}
{The Ada Lovelace Institute}: Technical methods for regulatory inspection of algorithmic systems (2021)

\bibitem[{Thelisson and Verma(2024)}]{thelisson_conformity_2024}
Thelisson, E., Verma, H.: Conformity assessment under the {EU} {AI} act general approach. AI and Ethics \textbf{4}(1), 113--121 (Feb 2024), ISSN 2730-5961, \urlprefix\url{https://doi.org/10.1007/s43681-023-00402-5}

\bibitem[{Thomas and Autio(2019)}]{thomas_innovation_2019}
Thomas, L.D.W., Autio, E.: Innovation {Ecosystems} (Oct 2019), \urlprefix\url{https://papers.ssrn.com/abstract=3476925}

\bibitem[{Trezza(2023)}]{trezza_scrape_2023}
Trezza, D.: To scrape or not to scrape, this is dilemma. {The} post-{API} scenario and implications on digital research. Frontiers in Sociology \textbf{8} (2023), ISSN 2297-7775, \urlprefix\url{https://www.frontiersin.org/articles/10.3389/fsoc.2023.1145038}

\bibitem[{Vakkuri et~al.(2020)Vakkuri, Kemell, Jantunen, and Abrahamsson}]{vakkuri_this_2020}
Vakkuri, V., Kemell, K.K., Jantunen, M., Abrahamsson, P.: “{This} is {Just} a {Prototype}”: {How} {Ethics} {Are} {Ignored} in {Software} {Startup}-{Like} {Environments}. In: Stray, V., Hoda, R., Paasivaara, M., Kruchten, P. (eds.) Agile {Processes} in {Software} {Engineering} and {Extreme} {Programming}, pp. 195--210, Lecture {Notes} in {Business} {Information} {Processing}, Springer International Publishing, Cham (2020), ISBN 978-3-030-49392-9

\bibitem[{Vakkuri et~al.(2022)Vakkuri, Kemell, Kultanen, Siponen, and Abrahamsson}]{vakkuri_ethically_2022}
Vakkuri, V., Kemell, K.K., Kultanen, J., Siponen, M., Abrahamsson, P.: Ethically {Aligned} {Design} of {Autonomous} {Systems}: {Industry} {Viewpoint} and an {Empirical} {Study}. EJBO Electronic Journal of Business Ethics and Organization Studies \textbf{27}(1) (2022)

\bibitem[{{van de Waerdt}(2020)}]{VANDEWAERDT2020105436}
{van de Waerdt}, P.J.: Information asymmetries: recognizing the limits of the gdpr on the data-driven market. Computer Law \& Security Review \textbf{38}, 105436 (2020), ISSN 0267-3649, \urlprefix\url{https://www.sciencedirect.com/science/article/pii/S0267364920300418}

\bibitem[{Veale(2019)}]{veale_governing_2019}
Veale, M.: Governing {Machine} {Learning} that {Matters}. Ph.D. thesis, University College London (UCL) (Aug 2019)

\bibitem[{Wehde(2022)}]{wehde-dsa-e}
Wehde: Datenzugang über art. 31 abs. 2 dsa-e. MMR pp. 827--830 (2022)

\bibitem[{Wieringa(2020)}]{wieringa_what_2020}
Wieringa, M.: What to account for when accounting for algorithms: a systematic literature review on algorithmic accountability. In: Proceedings of the 2020 {Conference} on {Fairness}, {Accountability}, and {Transparency}, pp. 1--18, {FAT}* '20, Association for Computing Machinery, New York, NY, USA (Jan 2020), ISBN 978-1-4503-6936-7, \urlprefix\url{https://dl.acm.org/doi/10.1145/3351095.3372833}

\bibitem[{Wilson et~al.(2021{\natexlab{a}})Wilson, Ghosh, Jiang, Mislove, Baker, Szary, Trindel, and Polli}]{10.1145/3442188.3445928}
Wilson, C., Ghosh, A., Jiang, S., Mislove, A., Baker, L., Szary, J., Trindel, K., Polli, F.: Building and auditing fair algorithms: A case study in candidate screening. In: Proceedings of the 2021 ACM Conference on Fairness, Accountability, and Transparency, p. 666–677, FAccT '21, Association for Computing Machinery, New York, NY, USA (2021{\natexlab{a}}), ISBN 9781450383097, \urlprefix\url{https://doi.org/10.1145/3442188.3445928}

\bibitem[{Wilson et~al.(2021{\natexlab{b}})Wilson, Ghosh, Jiang, Mislove, Baker, Szary, Trindel, and Polli}]{wilson_building_2021}
Wilson, C., Ghosh, A., Jiang, S., Mislove, A., Baker, L., Szary, J., Trindel, K., Polli, F.: Building and {Auditing} {Fair} {Algorithms}: {A} {Case} {Study} in {Candidate} {Screening}. In: Proceedings of the 2021 {ACM} {Conference} on {Fairness}, {Accountability}, and {Transparency}, pp. 666--677, ACM, Virtual Event Canada (Mar 2021{\natexlab{b}}), ISBN 978-1-4503-8309-7, \urlprefix\url{https://dl.acm.org/doi/10.1145/3442188.3445928}

\bibitem[{Wörsdörfer(2023)}]{Worsdorfer2023}
Wörsdörfer, M.: The e.u.'s artificial intelligence act: An ordoliberal assessment. AI Ethics  (2023)

\bibitem[{Young et~al.(2022)Young, Katell, and Krafft}]{10.1145/3531146.3533194}
Young, M., Katell, M., Krafft, P.: Confronting power and corporate capture at the facct conference. In: Proceedings of the 2022 ACM Conference on Fairness, Accountability, and Transparency, p. 1375–1386, FAccT '22, Association for Computing Machinery, New York, NY, USA (2022), ISBN 9781450393522, \urlprefix\url{https://doi.org/10.1145/3531146.3533194}

\end{thebibliography}

\end{document}